\documentclass[12pt]{article}
\pdfoutput=1
\usepackage[utf8]{inputenc}
\usepackage[DIV13]{typearea}
\usepackage{ragged2e}
\usepackage{calligra,amsmath,amsfonts,mathrsfs,amssymb,bbm,textcomp,bbding}
\usepackage{slashed,cancel,units}
\usepackage[usenames,dvipsnames]{xcolor}
\usepackage{cite}
\usepackage{hyperref}
\hypersetup{colorlinks=true,urlcolor=Magenta,anchorcolor=blue,citecolor=blue,filecolor=blue,
            linkcolor=Magenta,menucolor=blue, linktocpage=true,pdfproducer=medialab}
\usepackage[english]{babel}
\usepackage{catchfile}
\usepackage{indentfirst}
\usepackage{graphicx}
\usepackage{enumerate}
\newcommand{\email}[1]{\href{mailto:#1}{\tt #1}}

\numberwithin{equation}{section}

\newcommand{\getenv}[2][]{%
  \CatchFileEdef{\temp}{"|kpsewhich --var-value #2"}{}%
  \if\relax\detokenize{#1}\relax\temp\else\let#1\temp\fi}
\getenv[\USER]{USER}

\newcommand{\blue}[1]{\color{blue} #1 \color{black}}

\newcommand{\be}{\begin{equation}}
\newcommand{\ee}{\end{equation}}
\newcommand{\ba} {\begin{equation}\begin{aligned}}
\newcommand{\ea} {\end{aligned}\end{equation}}
\newcommand{\bea}{\begin{eqnarray}}
\newcommand{\eea}{\end{eqnarray}}
\newcommand{\nn}{\nonumber}

\newcommand{\paren}[1]{\left( #1 \right)}
\def\l{\left(}
\def\r{\right)}
\def\ov{\overline}
\def\diag{{\tt diag}}
\def\Tr{{\rm Tr}}
\def\hc{\mathrm{h.c.}}

\newcommand{\dmu}{\partial_\mu}
\newcommand{\dmup}{\partial^\mu}

\newcommand{\sq}{\sin\vartheta}
\newcommand{\cq}{\cos\vartheta}

\newcommand{\cmark}{\checkmark}
\newcommand{\xmark}{\texttimes}

\def\lsp{\lambda_{s\phi}}

\def\UPQ{U(1)_\text{PQ}}
\def\TeV{\text{ TeV}}
\def\GeV{\text{ GeV}}
\def\MeV{\text{ MeV}}
\def\keV{\text{ keV}}
\def\eV{\text{ eV}}
\def\meV{\text{ meV}}
\def\mueV{\text{ $\mu$eV}}
\def\m{\text{ m}}
\def\mm{\text{ mm}}

\newcommand{\LL}{\mathscr{L}}

\def\cO{\mathcal{O}}
\def\V{\mathbf{V}}
\newcommand{\M}{\mathcal{M}}
\newcommand{\Msq} {\mathcal{M}_f\mathcal{M}_f^\dag}

\def\lam{\lambda}
\def\Lam{\Lambda}

\def\s{\sigma}
\begin{document}
\renewcommand*{\thefootnote}{\fnsymbol{footnote}}
\begin{titlepage}
\vspace*{-1cm}
\phantom{hep-ph/***}
\flushleft{FTUAM-17-16}
\hfill{IFT-UAM/CSIC-17-082}
\hfill\\
\vskip 2cm
\begin{center}
\mathversion{bold}
\blue{{\LARGE\bf The Minimal Axion Minimal Linear $\sigma$ Model}}\\[4mm]
\mathversion{normal}
\vskip .3cm
\end{center}
\vskip 0.5  cm
\begin{center}
{\large\bf  L.~Merlo}~$^{a)}$\footnote{\email{luca.merlo@uam.es}},
{\large\bf F. Pobbe}~$^{b)}$\footnote{\email{federico.pobbe@pd.infn.it}}, and
{\large\bf S.~Rigolin}~$^{b)}$\footnote{\email{stefano.rigolin@pd.infn.it}}
\vskip .7cm
{\footnotesize
$^{a)}$~
Departamento de F\'isica Te\'orica and Instituto de F\'isica Te\'orica UAM/CSIC,\\
Universidad Aut\'onoma de Madrid, Cantoblanco, 28049, Madrid, Spain\\
\vskip .1cm
$^{b)}$~Dipartimento di Fisica e Astronomia, Universit\`a di Padova and\\ 
INFN, Sezione di Padova, via Marzolo 8, I-35131 Padova, Italy
}
\end{center}
\vskip 2cm
\begin{abstract}
\justify
The minimal $SO(5)/SO(4)$ linear $\sigma$ model is extended including an additional complex scalar field, singlet under the global $SO(5)$ and the Standard Model gauge symmetries. The presence of this scalar field creates the conditions to generate an axion {\it \`a la} KSVZ, providing a solution to the strong CP problem, or an axion-like-particle. Different choices for the PQ charges are possible and lead to physically distinct Lagrangians. The internal consistency of each model necessarily requires the study of the scalar potential describing the $SO(5)\to SO(4)$, electroweak and PQ symmetry breaking. A single minimal scenario is identified and the associated scalar potential is minimised including counter-terms needed to ensure one-loop renormalisability. In the allowed parameter space, phenomenological features of the scalar degrees of freedom, of the exotic fermions and of the axion are illustrated. Two distinct possibilities for the axion arise: either it is a QCD axion with an associated scale larger than $\sim10^{5}$ TeV and therefore falling in the category of the invisible axions; or it is a more massive axion-like-particle, such as a $1$ GeV axion with an associated scale of $\sim1$ TeV, that may show up in collider and flavour searches.
\end{abstract}
\end{titlepage}
\setcounter{footnote}{0}

\pdfbookmark[1]{Table of Contents}{tableofcontents}
\tableofcontents

\renewcommand*{\thefootnote}{\arabic{footnote}}
\section{Introduction}

The last decade experienced a revival of interest for the so-called Composite Higgs (CH) models: first introduced in the middle 
of the '80s~\cite{Kaplan:1983fs,Kaplan:1983sm,Banks:1984gj}, they have been reconsidered 20 years later with a more economical symmetry 
content~\cite{Agashe:2004rs,Gripaios:2009pe,Mrazek:2011iu}. The Minimal Composite Higgs Model (MCHM)~\cite{Agashe:2004rs} is based 
on the non-linear realisation of the $SO(5)/SO(4)$ spontaneous breaking, which relies on a not well identified strong dynamics: the 
four Nambu-Goldstone bosons (GBs), originated from the global symmetry breaking, can be identified with the three would-be longitudinal 
components of the Standard Model (SM) gauge bosons and the Higgs field. The gauging of the SM symmetry group and the interactions with the 
SM fermions produce an explicit mass term for the Higgs field, which otherwise would be massless due to the underlying GB shift symmetry. 
This mechanism provides an elegant solution to the so-called Electroweak (EW) Hierarchy Problem. 

A general drawback of these CH constructions is represented by its effective formulation: the generality of the effective approach 
comes together with its limited energy range of application. Refs.~\cite{Barbieri:2007bh,Gertov:2015xma,Feruglio:2016zvt,Gavela:2016vte} 
attempted to improve in this respect, providing a renormalisable description of the scalar sector. Following for definiteness 
the treatment done in Ref.~\cite{Feruglio:2016zvt}, the Minimal $SO(5)/SO(4)$ Linear $\sigma$ model (ML$\sigma$M) is constructed 
extending the SM spectrum by the introduction of an EW singlet scalar field $\sigma$ and a specific set of vector-like fermions 
in the singlet and in the fundamental representations of $SO(5)$. In the limit of large $\sigma$ mass, the model falls back 
onto the usual effective non-linear description of the MCHM~\cite{Agashe:2004rs,Barbieri:2007bh,Alonso:2014wta,Panico:2015jxa,
Hierro:2015nna}, that represents a specific realisation of the so-called Higgs Effective Field Theory~\cite{Feruglio:1992wf,
Grinstein:2007iv,Contino:2010mh,Alonso:2012px,Alonso:2012jc,Alonso:2012pz,Buchalla:2013rka,Brivio:2013pma,Brivio:2014pfa,
Gavela:2014vra,Gavela:2014uta,Alonso:2015fsp,Gavela:2016bzc,Alonso:2016btr,Eboli:2016kko,Brivio:2016fzo,Alonso:2016oah,
deFlorian:2016spz,Merlo:2016prs,Alonso:2017tdy,Buchalla:2017jlu,Kozow:2019txg} Lagrangian describing the most general Higgs couplings to SM 
gauge bosons and fermions, which preserve the SM gauge symmetry.

The ML$\sigma$M can also be considered an optimal framework where to look for a solution to the strong CP problem. Indeed, 
extending the scalar spectrum with an additional complex scalar field $s$, $SO(5)$ and EW singlet, the symmetry content of 
the model can be supplemented with an extra Peccei-Quinn (PQ) $\UPQ$~\cite{Peccei:1977hh}, eventually providing a realisation 
of the KSVZ axion mechanism~\cite{Kim:1979if,Shifman:1979if}: the angular component of the extra scalar $s$ may indeed represent 
an axion\footnote{In Ref.~\cite{Gripaios:2016mmi} the MCHM has been enriched by an additional $U(1)$ symmetry, that is non-anomalous 
and therefore does not originate a QCD axion.}. This idea has been firstly developed in Ref.~\cite{Brivio:2017sdm} and this class 
of models will be dubbed Axion Minimal Linear $\sigma$ Model (AML$\s$M). Even in this simple setup, the choice of the PQ charge 
assignment is not unique and different choices lead to physically distinct Lagrangians. 

In this paper, a ``minimality criterium'' in terms of number of parameters will be introduced and only one ``minimal scenario'', 
the minimal AML$\s$M, is identified among all the constructions presented in Ref.~\cite{Brivio:2017sdm}. In order to completely fix 
the PQ charge assignment the following requirements are imposed: the SM fermion masses are generated at tree-level through the fermion 
partial compositeness mechanism~\cite{Kaplan:1991dc,Contino:2004vy,Dugan:1984hq,Galloway:2010bp}, which is the only explicit $SO(5)$ 
breaking sector; the PQ scalar field $s$ couples to (part of) the exotic fermions providing a portal between the axion and the colour 
interactions. The angular component of $s$ can be identified as a QCD axion, requiring in addition that the contributions to the 
colour anomaly allow to reabsorb the QCD-$\theta$ parameter through a shift symmetry transformation, thus solving the strong CP problem. 
If instead this requirement is relaxed, then the PQ GB si dubbed axion-like-particle (ALP). Both the possibilities are envisaged in 
the minimal AML$\s$M identified through the conditions aforementioned. Moreover, in this scenario, all the SM fields do not transform 
under the PQ symmetry and three distinct scales are present, that is the EW scale, the $SO(5)/SO(4)$ and PQ symmetry breaking scales, 
the latter being independent from the first two.

A dedicated analysis of the scalar potential and its minima is necessary in order to guarantee that $SO(5)$ gets spontaneously broken 
down to $SO(4)$, and that the EW symmetry breaking (EWSB) mechanism occurs providing the correct EW vacuum expectation value (VEV). 
This analysis requires to take into account contributions to the scalar potential arising at one one-loop from the fermions and the 
gauge bosons of the model. The renormalisable scalar potential is identified according to the aforementioned requirements. The 
associated parameter space is studied, both analytically for few limiting cases and numerically, illustrating the main features of 
this minimal model. The phenomenological analysis reveals that modifications of the Higgs couplings to SM fermions and gauge 
bosons are present, leading to possibly interesting signals at colliders. 

Turning the attention to the PQ GB sector, the axion and the ALP cases are characterised by two distinct phenomenologies. The axion is very 
light, with a mass generated by non-perturbative QCD effects as in the traditional PQ models~\cite{Peccei:1977hh,Wilczek:1977pj,
Weinberg:1977ma,Bardeen:1978nq,DiVecchia:1980yfw}. Its corresponding scale is larger than $\sim10^5\TeV$ and therefore it enters 
into the category of the invisible axion models~\cite{Kim:1979if,Shifman:1979if,Dine:1981rt,Zhitnitsky:1980tq}. On the other side, 
the ALP can be much heavier, but at the price of invoking a soft explicit breaking of the shift symmetry and not necessarily solving 
the strong CP problem. As its characteristic scale can be much lower, it may give rise to visible effects at colliders. 

It is the aim of the present paper to illustrate in details the minimal AML$\s$M and to analyse its phenomenological features. 
In the next section, the construction of the AML$\sigma$M is described, discussing the fermion content and the main characteristics 
of the scalar potential, focussing on the renormalisability of the full Lagrangian. In Sect.~\ref{Sect.PQCharges}, the minimal 
scenario is identified, based on a minimality criterium in terms of number of parameters of the whole Lagrangian. 
Sect.~\ref{Sect.ScalarPotential} is devoted to the analytical description of the scalar potential and the $SO(5)/SO(4)$ spontaneous 
symmetry breaking mechanism, presenting few relevant limiting cases. The phenomenological features of the model are described in 
Sect.~\ref{Sect.Pheno} and Sect.~\ref{Sect.AxionPhen}, with the latter section dedicated to the analysis of the axion and of the ALP. 
Finally, conclusions are drawn in Sect.~\ref{Sect.Conclusions}, while more technical details are left for the appendix.

\boldmath
\section{The Axion Minimal Linear $\sigma$ Model}
\label{Sect.AxionMLsModel}
\unboldmath

The ML$\s$M based on the linear $SO(5)/SO(4)$ symmetry breaking realisation has been analysed in Ref.~\cite{Feruglio:2016zvt}. 
As usual in this class of minimal models, an additional $U(1)_X$ is introduced in order to ensure the correct hypercharge 
assignment. The field content of the original ML$\sigma$M is the following:
\begin{enumerate} 
\item The four SM gauge bosons associated to the SM gauge symmetry.
\item A real scalar field $\phi$ in the fundamental representation of $SO(5)$, which includes the three would-be-longitudinal 
components of the SM gauge bosons $\pi_i$, $i=1,\,2,\,3$, the Higgs field $h$ and the additional complex scalar field $\s$, 
singlet under the SM gauge group:
\be
\phi=\left(\pi_1,\,\pi_2,\,\pi_3,\,h,\,\sigma\right)^T
\qquad
\xrightarrow{u.g.}
\qquad
\phi=\left(0,\,0,\,0,\,h,\,\sigma\right)^T\,,
\label{phiDefinition}
\ee
where the last expression holds when selecting the unitary gauge, which will be used throughout the next sections.

\item Exotic vector fermions, which couple directly to the $SO(5)$ scalar sector through $SO(5)$ invariant proto--Yukawa 
interactions. These fermions transform either in the fundamental of $SO(5)$, and they will be labelled as $\psi$, or in the 
singlet representation of $SO(5)$, dubbed $\chi$. For both types of fermions, two distinct $U(1)_X$ assignments 
are considered, $2/3$ and $-1/3$, as they are necessary to induce mass terms for both the SM up and the down quark sectors.

\item SM fermions, which do not couple directly to the Higgs field. SM fermion masses are originated through SM--exotic 
fermion interactions in the spirit of the fermion partial compositeness mechanism~\cite{Kaplan:1991dc,Contino:2004vy,
Dugan:1984hq,Galloway:2010bp}. SM fermions do not come embedded in a complete representation of $SO(5)$, leading to a soft explicit 
$SO(5)$ symmetry breaking. Although the whole SM fermion sector could be considered, only the top and bottom quarks will be retained 
in what follows. This simplification does not have relevant consequences on the results presented here and the three generation setup 
can be easily envisaged.
\end{enumerate}

The AML$\sigma$M encompasses, in addition to the previous content, 
\begin{enumerate}
\setcounter{enumi}{4}
\item A complex scalar field $s$, singlet under the global $SO(5) \times U(1)_X$ and the SM gauge group. Adopting an 
exponential notation,
\be
s \equiv\dfrac{r}{\sqrt2}e^{ia/f_a}\,,
\label{sDefinition}
\ee 
the degrees of freedom are defined as the radial component $r$ and the angular one $a$, to be later identified with the 
physical axion. Following the philosophy adopted in Ref.~\cite{Feruglio:2016zvt} any direct coupling 
between the scalar $s$ and the SM fermions is not introduced, as it will be discussed in more details in the following.
\end{enumerate}

The complete renormalisable Lagrangian for the AML$\sigma$M can be written as the sum of three terms describing respectively 
the pure gauge, fermionic and scalar sectors, 
\be
\LL=\LL_{\rm g} + \LL_{\rm f}+ \LL_{\rm s}\,.
\label{FullLag}
\ee
The explicit expression for each piece will be detailed in the following subsections.

\subsection{The Gauge Lagrangian}

The first term, $\LL_{\rm g}$, contains the SM gauge kinetic and the colour anomaly terms,
\be
\LL_{\rm g}=-\dfrac{1}{4}G^{a\mu\nu}G^a_{\mu\nu}-\dfrac{1}{4}W^{a\mu\nu}W^a_{\mu\nu}-\dfrac{1}{4}B^{\mu\nu}B_{\mu\nu}+
            \dfrac{\alpha_s}{8\pi}\theta G^{a\mu\nu}\widetilde{G}^a_{\mu\nu}\,,
\ee 
with the indices summed over $SU(3)_c$ or $SU(2)_L$, and
\be
\widetilde{G}_{\mu\nu}\equiv\dfrac{1}{2}\epsilon_{\mu\nu\rho\sigma}G^{\rho\sigma} \qquad \qquad ({\rm with} \quad 
          \epsilon_{1230}=+1) \,.
\ee
The introduction of the axion will provide a natural explanation for the vanishing of the QCD-$\theta$ term.

\subsection{The Fermionic Lagrangian}

According to the spectrum and symmetries described in the previous section, the fermionic part of the renormalisable 
Lagrangian in agreement with Ref.~\cite{Brivio:2017sdm}, although with a slightly different notation, reads
\be
\begin{aligned}
\LL_{\rm f} =&\,\, 
\overline{q}_L i\slashed{D}\,q_L\,+\,\overline{t}_R i\slashed{D}\,t_R\,+\,
                                     \overline{b}_R i\slashed{D}\,b_R \,+\,  \\
   &+ \overline{\psi} \left[i \slashed{D} - M_5 \right] \psi \,+\, 
       \overline{\chi} \left[i \slashed{D} - M_1 \right]\chi \,-\,
       \left[ y_1\, \overline{\psi}_L\, \phi\, \chi_R + y_2\, \overline{\psi}_R\, \phi\, \chi_L +\hc \right]+  \\
   &- \left[z_1\,\overline{\chi}_R\,\chi_L\,s + \tilde{z}_1\,\overline{\chi}_R\,\chi_L\,s^\ast + 
      z_5\,\overline{\psi}_R\,\psi_L\,s + \tilde{z}_5\,\overline{\psi}_R\,\psi_L\,s^\ast + \hc \right] \,+  \\
   &+ \left[\Lambda_1 \left(\overline{q}_L \Delta_{2\times5} \right) \psi_R  
    +  \Lambda_2 \, \overline{\psi}_L \left (\Delta_{5\times1} t_R \right) 
    +  \Lambda_3 \,\overline{\chi}_L t_R + \hc \right]\,+ \\
   &+ \overline{\psi'} \left[i \slashed{D} - M'_5 \right]\psi' \,+\, 
       \overline{\chi'} \left[i \slashed{D} - M'_1 \right]\chi' \,-\,
       \left[y'_1\, \overline{\psi'}_L\, \phi\, \chi'_R + y'_2\,\overline{\psi'}_R\, \phi\,\chi'_L +\hc\right]+  \\
   &- \left[z'_1\,\overline{\chi'}_R\,\chi'_L\,s + \tilde{z}'_1\,\overline{\chi'}_R\,\chi'_L\,s^\ast + 
      z'_5\,\overline{\psi'}_R\,\psi'_L\,s + \tilde{z}'_5\,\overline{\psi'}_R\,\psi'_L\,s^\ast + \hc \right]+ \\ 
   &+ \left[\Lambda'_1 \left(\overline{q}_L \Delta'_{2\times5} \right) \psi'_R  
    +  \Lambda'_2 \,\overline{\psi'}_L \left (\Delta'_{5\times1} b_R \right) 
    +  \Lambda'_3 \,\overline{\chi'}_L b_R + \hc \right]\,. 
\end{aligned}
\label{FermionLagrangian}
\ee
The first line contains the kinetic terms for the 3$^{rd}$ generation SM quarks, being $q_L$ the left-handed (LH) $SU(2)_L$ 
doublet and $t_R$ and $b_R$ the right-handed (RH) singlet counterparts. The second line contains the kinetic and mass terms 
for the exotic vector fermions, $\psi$ and $\chi$ (with $U(1)_X$ charge $2/3$). The direct mass terms for the heavy fermions 
are denoted by $M_{1,5}$ respectively for the fermions in the singlet and fundamental representations. The proto-Yukawa couplings 
between the heavy fermions and the real scalar quintuplet field $\phi$ are also present in the second line. In the third line, 
the Yukawa-like couplings of the exotic fermions with the complex scalar singlet $s$ are shown. Two distinct type of couplings, 
$z$ and $\tilde{z}$, have been introduced reflecting the freedom in choosing the PQ charges of $s$ and of the fermionic bilinears. 
The fourth line contains the interactions between the top quark and exotic fermions with $U(1)_X$ charge equal to $2/3$. 

While, the second and third lines of the Lagrangian explicitly preserve $SO(5)$, the partial compositeness terms in the fourth line,  
proportional to $\Lambda_{1,2}$, explicitly break the global $SO(5)$ symmetry. The combinations $\Lambda_1\Delta_{2\times 5}$ and 
$\Lambda_2\Delta_{5\times 1}$ may play the role of spurions~\cite{DAmbrosio:2002vsn,Cirigliano:2005ck,Davidson:2006bd,Grinstein:2010ve,Feldmann:2010yp,Guadagnoli:2011id,Alonso:2011jd,Buras:2011zb,Buras:2011wi,Lopez-Honorez:2013wla,Alonso:2016onw,Dinh:2017smk,Merlo:2018rin} that formally ensure the $SO(5)\times U(1)_X$ invariance of the operators~\footnote{Steps forward the spurion approach consist in promoting the spurions to   dynamical scalar fields and studying the corresponding scalar potential~\cite{Alonso:2011yg,Alonso:2012fy,Alonso:2013mca,Alonso:2013nca} (see also Refs.~\cite{Anselm:1996jm,Barbieri:1999km,Berezhiani:2001mh,Feldmann:2009dc,Nardi:2011st}).}. The exotic fermion spinors can be decomposed 
under the $SU(2)_L$ quantum numbers as follows:
\be
\psi\sim\left(K,\,Q,\,T_5\right)\,,\qquad\qquad \chi\sim T_1\,,
\label{PsiChiComponents}
\ee
being $K$ and $Q$ doublets while $T_{1,5}$ singlets of $SU(2)_L$. The resulting interactions preserve the gauge EW 
symmetry, with the hypercharge defined as
\be
Y=\Sigma_R^{(3)}+X \,,
\ee
with $\Sigma_R^{(3)}$ the third component of the global $SU(2)_R$ ($1/2$ for $K$ and $-1/2$ for $Q$) and $X$ the $U(1)_X$ 
charge of the spinor.

The last three lines describe the replicated sector associated to the bottom quark. The exotic vector fermions, $\psi'$ and 
$\chi'$ have $U(1)_X$ charge $-1/3$ to allow the direct partial compositness coupling with the bottom. Their decomposition 
in terms of $SU(2)_L$ representations, reads
\be
\psi'\sim\left(Q',\,K',\,B'_5\right)\,,\qquad\qquad \chi'\sim B'_1\,,
\label{PsiChiComponents}
\ee
being $Q'$ and $K'$ doublets of $SU(2)_L$ (with $\Sigma_R^{(3)}$ component $1/2$ and $-1/2$ respectively) and $B'_{1,5}$ 
singlets of $SU(2)_L$. 

The Lagrangian in Eq.~(\ref{FermionLagrangian}) can be rewritten for later convenience in terms of fermionic vectors 
regrouping all the spinors components ordered accordingly of their electric charge,
\be
\Psi=\left(K^u,\,\mathcal{T},\,\mathcal{B}\,,K^{\prime d}\right)\,,
\ee
with
\be
\mathcal{T}=\left(t,\, Q^u,\,K^d,\,T_5,\, T_1,\,Q^{\prime u}\right)\,,\qquad\qquad
\mathcal{B}=\left(b,\, Q^{\prime d},\,K^{\prime u},\,B'_5,\, B'_1,\,Q^d\right)\,.
\ee
The fermion mass terms in Eq.~(\ref{FermionLagrangian}) can then be written as
\be
\LL_\mathcal{M}=-\ov\Psi_L\,\mathcal{M}_f(h,\sigma,r)\,\Psi_R\,,
\ee
where the field dependent fermion mass matrix $\mathcal{M}_f$ is a $14\times14$ block diagonal matrix,
\be
\mathcal{M}_f(h,\sigma,r)=\diag\Big(M_5(r),\,\mathcal{M}_\mathcal{T}(h,\sigma,r),\,\mathcal{M}_\mathcal{B}(h,\sigma,r),\,M'_5(r)\Big)\,.
\label{BigMassFermion}
\ee
For the top sector one has explicitly 
\bea
\mathcal{M}_\mathcal{T}(h,\sigma,r)&=&
\left(
\begin{array}{cccccc}
0 	& \Lambda_1 	& 0 		& 0 		& 0 			& \Lambda'_1\\
0 	& M_5(r) 	& 0 		& 0 		& y_1\frac{h}{\sqrt2} 	& 0\\
0 	& 0		& M_5(r) 	& 0		& y_1\frac{h}{\sqrt2}	& 0\\
\Lambda_2	& 0	& 		& M_5(r)	& y_1 \sigma		& 0\\
\Lambda_3	& y_2\frac{h}{\sqrt2}	& y_2\frac{h}{\sqrt2}	& y_2\sigma	& M_1(r)	& 0\\
0	& 0		& 0		& 0		& 0			& M'_5(r)
\end{array}
\right) \, ,
\label{FermionMassMatrixComponents}
\eea
with
\be
M_1(r)=M_1+\left(z_1+\tilde{z}_1\right) r \,, \qquad\qquad
M_5(r)=M_5+\left(z_5+\tilde{z}_5\right) r\,.
\label{rmasses}
\ee
The corresponding matrix for the bottom sector, $\mathcal{M}_\mathcal{B}(h,\sigma,r)$ can be obtained from 
Eqs.~(\ref{FermionMassMatrixComponents}) and (\ref{rmasses}) by replacing the unprimed couplings with the corresponding 
primed ones.

Eqs.~(\ref{FermionLagrangian}), (\ref{FermionMassMatrixComponents}), and (\ref{rmasses}) contain all the possible 
couplings invariant under SM gauge group and $SO(5)\times U(1)_X$ 
global symmetry that can be constructed following the assumptions described in the previous section. However, it is 
important to notice that the Lagrangian actually describing the AML$\s$M can be obtained only after the adoption of a 
specific choice for the PQ charges: not all the terms are simultaneously allowed. In fact, only one between the 
$M_i$, $z_i$ and $\tilde{z}_i$ (and corresponding primed) terms is allowed once a specific PQ charge assignment for 
the fermion chiral components is chosen, assuming obviously a non--vanishing charge for the scalar $s$ field. 
In other words, exotic fermions acquire masses either through the direct mass terms ($M_i$) or through the Yukawa--like 
interactions with $s$ ($z_i$ or $\tilde{z}_i$) once the scalar field $s$ develops a VEV. In addition, following the 
assumptions outlined in the previos section, as the scalar quintuplet $\phi$ does not transform under the PQ symmetry, 
the presence of the proto--Yukawa interactions ($y_{i}$) necessarily depend on the PQ charges of exotic fermions.

Finally, turning the attention to the interactions between exotic and SM fermions, in the fourth and seventh 
lines of Eq.~(\ref{FermionLagrangian}), if only the exotic fermions have non-vanishing PQ charges, then these operators 
are forbidden, unless the $\Lambda_{i}$ couplings are either promoted to be spurions under the PQ symmetry or substituted 
by a PQ dynamical field ($s$ or $s^*$). This would introduce explicit sources for the PQ symmetry breaking or imply that 
the PQ sector contributes to the dynamics that originate these operators. These issues will be discussed in the next 
sections, where the conditions that lead to the minimal AML$\s$M charge assignment are illustrated.

\subsection{The Scalar Lagrangian}

The scalar part of the Lagrangian introduced in Eq.~(\ref{FullLag}) describes scalar-gauge and scalar-scalar interactions:
\be
\LL_{\rm s} = \frac{1}{2}(D_\mu\phi)^T (D^\mu\phi) + (\dmu s^*) (\dmup s) - V(\phi,s)\,,
\label{Lags}
\ee
where the $SU(2)_L\times U(1)_Y$ covariant derivative of the quintuple $\phi$ is given by
\be
D_\mu\phi=\paren{\dmu+ i g \Sigma_L^{(i)} W_\mu^i+ig'\Sigma_R^{(3)}B_\mu}\phi \,,
\label{covariant}
\ee 
and $\Sigma_L^i$ and $\Sigma_R^i$ denote respectively the generators of the $SU(2)_L \times SU(2)_R \sim SO(4)'$ subgroup of 
$SO(5)$, rotated with respect to the $SO(4)$ group preserved from the spontaneous breaking. 

It will be useful for later convenience to express the scalar Lagrangian in Eq.~(\ref{Lags}) in the unitary gauge, making 
use of Eqs.~(\ref{phiDefinition}) and (\ref{sDefinition}):
\be
\begin{split}
\LL_{\rm s} =& \phantom{+}\frac{1}{2}(\dmu h)(\dmup h) + \frac{1}{2}(\dmu \s)(\dmup \s) \,+\, 
            \frac{h^2}{4} \left[ g^2\,W^+_\mu W^{-\mu} + \frac{g^2+{g'}^2}{2} Z_\mu Z^\mu \right] \,+ \\
             &+ \frac{1}{2}(\dmu r)(\dmup r) + \frac{r^2}{2 f_a^2}(\dmu a)(\dmup a) - V(h,\s,r) \,,
\end{split}
\label{Lagr}             
\ee

Notice that once the $\UPQ$ 
gets spontaneously broken through the VEV of $r$, the kinetic term of the axion field $a$ gets canonically normalised, by identifying 
\be
f_a\equiv v_r\,.
\label{favsvr}
\ee

The scalar potential $V(\phi,s)$ can then be written as:
\be
V(\phi,s) =  V^\text{SSB}(\phi, s) +  V^\text{CW}(\phi, s) + V^\text{c.t.}(\phi, s)\,.
\ee
The first part, $V^\text{SSB}(\phi, s)$, describes the most general potential constructed out of $\phi$ and $s$, invariant under 
$SO(5)\times \UPQ$ symmetry, broken spontaneously to $SO(4)$:
\be
V^\text{SSB}(\phi,s) = \lambda (\phi^T\phi-f^2)^2 + \lambda_s (2\,s^*s-f_s^2)^2 - 2\lsp\,(s^*s)\,\left(\phi^T\phi\right) \,,
\label{VSSB}
\ee
where $\lambda$, $\lambda_s$ and $\lsp$ are the dimensionless quartic coefficients and the sign in front of $\lsp$ has been chosen 
negative for future convenience. Notice that $\lsp$ plays the role of portal between the $SO(5)$ and the PQ sectors: if $\lsp\sim\cO(1)$ 
then the $SO(5)/SO(4)$ and PQ breaking mechanisms would be linked and they would occur at similar scales; this would represent 
a possible tension between the naturalness of the AML$\s$M, which requires $f$ not so much larger than EW scale $v=246\GeV$, 
in order to reduce the typical fine-tuning in CH models, and the experimental data on the axion sector, which suggests very high 
values of $f_s$ (see Sect.~\ref{Sect.AxionPhen}). In consequence, values of $\lsp$ smaller than 1 are favoured in the AML$\s$M.

The expression of $V^\text{SSB}$ in the exponential notation will be useful in the following sections:
\be
V^\text{SSB}(h,\sigma,r) = \lambda(h^2+\sigma^2-f^2)^2 + \lambda_s (r^2-f_s^2)^2 - \lsp\,r^2\, (h^2+\sigma^2)  \,.
\label{VSSBUnitary}
\ee
When the scalar fields $h$, $\s$ and $r$ take a non trivial VEV, repectively $v_h$, $v_\s$ and $v_r$, a spontaneous symmetry
breaking for the EW, the global $SO(5)$ and the PQ symmetry, is obtained.

The second term $V^\text{CW}(\phi, s)$ is the Coleman--Weinberg (CW) one--loop potential that provides an explicit and dynamical 
breaking of the original symmetries. Its form depends on the explicit structure of the fermionic and bosonic Lagrangians and it 
will be outlined in the following subsection. 

Finally, the term $V^\text{c.t.}(\phi, s)$, includes all the couplings that need to be introduced at tree--level in order to 
cancel the divergences potentially arising from the one--loop CW potential, so to make the theory renormalizable.

\boldmath
\subsubsection*{The Coleman-Weinberg One-Loop Potential}
\unboldmath

Explicit dynamical breaking of the tree-level symmetries can be introduced at one-loop level through the CW mechanism~\cite{Coleman:1973jx}. 
Indeed, the presence of $SO(5)$ breaking couplings in both the fermionic and the gauge sectors generate $SO(5)$ breaking terms at 
one-loop level. Explicit $\UPQ$ breaking contributions may also be generated, depending on the fermion PQ charge assignment.  

The one-loop fermionic contributions can be calculated from the field dependent fermion mass matrix $\mathcal{M}_f(h,\,\s,\,r)$ 
in Eq.~(\ref{BigMassFermion}), using the usual CW expression:
\bea
V_f^\text{CW} &=& -\frac{1}{64\pi^2} \l\Tr\left[\Msq\right]\Lambda^2-\Tr\left[\left(\Msq\right)^2\right]\log\l\frac{\Lambda^2}{\mu^2}\r\right.\,
                   + \nn \\
 & &\hspace{1.0cm} \left. + \Tr\left[\left(\Msq\right)^2 \log\l \frac{\Msq}{\mu^2} \r\right]-\frac{1}{2} \Tr\left[\left(\Msq\right)^2\right]\r\,,
\label{Vcwfull}
\eea
where $\Lambda$ is the ultraviolet (UV) cutoff scale while $\mu$ is a generic renormalisation scale. The two terms in the first line 
are divergent, quadratically and logarithmically respectively, while those in the second line are finite. For the model under discussion 
the possible divergent contributions read
\bea
\Tr\left[\Msq\right] &=& c_0 +  c_1 (s^\ast s) + c_2 \, (\phi^T \phi)\,,  \label{tracemm} \\
\Tr\left[\left(\Msq\right)^2\right] &=& d_0 + d_1\,(s^\ast s) + d_2\,(\phi^T\phi) + d_3\,(s^\ast s)^2 + d_4\,(\phi^T \phi)^2 + 
                                        d_5\,(\phi^T \phi)(s^\ast s)+\nn\\
             & & \hspace{-2.0cm} + \, \tilde{d}_1\s + \tilde{d}_2 h^2 + \hat{d}_1\s (s+s^\ast) + \hat{d}_2 (\phi^T\phi)(s+s^\ast) 
                                         + \hat{d}_3 (\phi^T\phi) (s s+s^\ast s^\ast)\,.
\label{tracemm2}
\eea
The terms in Eq.~(\ref{tracemm})  are already present in the tree level potential $V^\text{SSB}$ in Eq.~(\ref{VSSBUnitary}) and therefore 
the quadratic divergences can be absorbed by a redefinition of the initial Lagrangian parameters. This is not the case for the logarithmic 
divergent term that contains five new couplings, denoted with $\tilde{d}_{1,2}$ and $\hat{d}_{1,2,3}$ in Eq.~(\ref{tracemm2}). The ones 
proportional to $\tilde{d}_{1,2}$ and $\hat{d}_{1}$ are $SO(5)$ breaking terms, while the ones proportional to $\hat{d}_{2,3}$ are $SO(5)$ 
preserving. On the other side, $\hat{d}_{1,2,3}$ terms also explicitly break the PQ symmetry. If in a specific setup these terms were not 
vanishing, renormalisability of the model would then require the introduction of the corresponding structures in the tree-level potential.

The expressions for the top sector CW coefficients that provide an explicit breaking of the $SO(5)$ and/or of the PQ symmetries read:

\be
\begin{aligned}
\tilde{d}_1 &= 4 (y_1 M_1 + y_2 M_5) \Lambda_2 \Lambda_3 \\
\tilde{d}_2 &= y_2^2 \Lambda_1^2 - 2\,y_1^2 \Lambda_2^2 \\
\hat{d}_1 &=  2 \,y_1 (z_1 + \tilde{z}_1) \Lambda_2 \Lambda_3 + 2 \,y_2 (z_5 + \tilde{z}_5) \Lambda_2 \Lambda_3 \\
\hat{d}_2 &= 2\, y_1 y_2 (z_1 + \tilde{z}_1)  M_5 + 2\, y_1 y_2 (z_5 + \tilde{z}_5) M_1\\
\hat{d}_3 &= 2 \, y_1 y_2\,( z_1 z_5 \, +\, \tilde{z}_1 \tilde{z}_5)\,.
\end{aligned}
\label{fbreaking}
\ee
Similar contributions for the bottom sector are obtained by substituting the unprimed couplings in Eq.(\ref{fbreaking}) with the 
corresponding primed ones. As stated before, once a specific PQ charge assignment is assumed, some of the couplings in the Lagrangian 
are forbidden, and consequently the corresponding CW coefficients vanish, as it will be explicitly discussed in the next section. 

In a similar way the one-loop gauge boson contributions to the CW potential can be calculated through the CW formula given in 
Eq.~(\ref{Vcwfull}) just substituting the fermion mass matrix $\M_f$ with the gauge boson one $\M_g$:
\be
V_g^\text{CW}= -\frac{1}{64\pi^2} \left(\Tr\left[\M_g^2\right]\Lambda^2 - 
                \Tr\left[\left(\M_g^2\right)^2 \right]\log\l\frac{\Lambda^2}{\mu^2}\r\,+ \ldots\right)\,. 
\label{tracemm2g}
\ee
The quadratic and logarithmic divergent terms read
\be
\Tr\left[\M_g^2\right] = \tilde{a}_1 h^2 \qquad\qquad\qquad
\Tr\left[\left(\M_g^2\right)^2\right] = b_0 + \tilde{b}_1\,h^4 \,,
\ee
with 
\be
\tilde{a}_1 = \dfrac{1}{8} \left(g^2 + g^{\prime2}\right)\qquad\qquad\qquad
\tilde{b}_1 = \dfrac{1}{64}\left[2\, g^4 + \left(g^2 + g^{\prime2}\right)^2\right]\,, 
\label{gbreaking}
\ee
both explicitly breaking the global $SO(5)$ symmetry. 

The two divergences associated to $\tilde{a}_1$ and $\tilde{d}_2$ require the introduction of an $h^2$ term in the tree--level scalar 
potential, in order to ensure the renormalisability of the model, while the divergence proportional to the $\tilde{b}_1$ coefficient 
requires an additional $h^4$ term.

\section{The Minimal Model}
\label{Sect.PQCharges}

There is a large zoology of possible $\UPQ$ charges that can be assigned to the spectrum discussed in the previous sections (see 
Ref.~\cite{Brivio:2017sdm} for details on more general charge assignments). However, after requiring a few, strong physical conditions, 
only one single set of charge assignments can be identified, which lead to the identification of the minimal AML$\s$M. The requirements 
are the following:
\begin{enumerate}
\item Mass terms for the SM quarks are originated at tree-level. Generalising the result in Ref.~\cite{Feruglio:2016zvt}, the leading 
order (LO) contribution to the third generation quark masses is given by
\be
m_t=\dfrac{y_1\Lambda_1\Lambda_3 v_h}{M_1(v_r)M_5(v_r)-y_1 y_2(v_h^2+v_\sigma^2)} - 
      \dfrac{y_1 y_2 \Lambda_1\Lambda_2v_h v_\sigma}{M_1(v_r)M_5^2(v_r)-y_1y_2M_5(v_r)(v_h^2+v_\sigma^2)}\,,
\label{TopQuarkMass}
\ee 
and similarly for the bottom mass. In this expression, $M_{1,5}(v_r)$ refer to the definitions in Eq.~(\ref{FermionMassMatrixComponents}) 
substituting the dependence on $r$ with its VEV, $v_r$. In order for this expression not to be vanishing, the conditions $y_1\neq0$ 
and $\Lambda_1\neq 0$ should hold simultaneously. Then, either $\Lambda_3\neq 0$ or $y_2\neq0 \wedge \Lambda_2\neq0$ should be verified, 
depending on whether the leading or sub-leading term in the $v/M$ expansion is retained.
\item The dynamics that generate the partial-composite operators in the fourth line of Eq.~(\ref{FermionLagrangian}) are associated 
only to the $SO(5)/SO(4)$ breaking sector. This implies that the scales $f$ and $f_s$ are distinct and independent. 
\end{enumerate}

In a completely generic model a third condition can be also considered:
\begin{enumerate}
\item[3.]
No PQ explicit breaking is generated at one-loop from the CW potential\footnote{The discussion on the consequences of PQ explicit breaking 
contributions, on its interest in cosmological studies, and on the case where the $SO(5)/SO(4)$ and PQ symmetry breaking occur at the 
same scale is deferred to Ref.~\cite{USFUTURE}.}. This condition is satisfied by imposing $\hat{d}_i=0$, for $i=1,\,2,\,3$ (and the 
equivalent ones for the bottom sector).
\end{enumerate}
This condition prevents the arising of large contributions to the axion mass, and it is automatically verified in the class of AML$\s$M 
constructions defined in Eq.~(\ref{FermionLagrangian}), as it will be explicitly shown in the following.

If one requires additionally to solve the strong CP problem {\it \`a la} KSVZ a fourth condition is necessary:
\begin{enumerate}
\item[4.] The complex scalar field $s$ needs to couple to at least one of the exotic fermions (not necessarily to all of them) and the 
net contribution to the QCD-$\theta$ term of the colour anomaly needs to be non-vanishing.
\end{enumerate}
This last condition, when satisfied, implies condition 3 and therefore for a QCD axion no PQ explicit breaking contributions arise in 
the scalar potential, besides those due to non-perturbative QCD effects.

The model identified with the PQ charge assignments in Tab.~\ref{tab:charges} satisfies to all the previous conditions: using the 
freedom to fix one of the charges, i.e. the charge of the complex scalar singlet $n_s=1$, the two cases shown in the table 
are physically equivalent.  This model is contained within the classes of constructions recently presented in Ref.~\cite{Brivio:2017sdm}.

\begin{table}[h!]
\centering{
\begin{tabular}{|c|c||c|c|c|c||c|c|c|c|c|c|c|c|c|}
\hline
$n_{q_L}$ & $n_{t_R}$ & $n_{\psi_L}$ & $n_{\psi_R}$ & $n_{\chi_L}$ & $n_{\chi_R}$ & $y_1$ & $y_2$ & $\Lambda_1$ & 
$\Lambda_2$ & $\Lambda_3$ & $M_5$ &$ M_1$ & $z_1,\tilde{z}_5$ & $\tilde{z}_1, z_5$ \\[1mm]
\hline
0 & 0 & $+1$ & 0 & 0 & $+1$ & \cmark & \cmark & \cmark & \xmark & \cmark & \xmark & \xmark & \cmark & \xmark  \\ 
0 & 0 & $-1$ & 0 & 0 & $-1$ & \cmark & \cmark & \cmark & \xmark & \cmark & \xmark & \xmark & \xmark & \cmark \\
\hline
\end{tabular}
\caption{\em On the left-side, the PQ charge assignments where $n_i$ refers to the $i$ field, 
conventionally fixing the PQ charge of the complex scalar field $s$, $n_s=1$. On the right-side, the parameters entering the 
fermionic Lagrangian, together with the information on whether they are compatible (\cmark) or not (\xmark) with the PQ symmetry. 
This assignment can be trivially extended to the bottom sector.}
\label{tab:charges}
}
\end{table}

The model presents a series of interesting features. No PQ charge is assigned to the SM particles and neither 
to the exotic fermions $\psi_R$ and $\chi_L$. The Yukawa-like terms proportional to $y_{1,2}$ are invariant under $\UPQ$, while 
the term proportional to $\Lambda_2$ is not and then it cannot be introduced in the Lagrangian. In consequence, the subleading 
contribution to the SM fermion masses is identically vanishing and the top mass is given only by the leading term in 
Eq.~(\ref{TopQuarkMass}) (similarly for the bottom mass).  The Dirac mass terms $M_{1,5}$ are also forbidden and then the exotic 
fermions $\psi$ and $\chi$ receive mass of the order $z_5 v_r$ (or $\tilde{z}_5 v_r$ depending on the specific sign of the PQ charge) 
and $z_1 v_r$ (or $\tilde{z}_1 v_r$), once $r$ develops a non-vanishing VEV. As $v_r$ is typically expected to be of the order of $f_s$, 
these fermions decouple from the spectrum when $f_s \gg f$. Finally, condition 2 implies that the couplings $\Lam_{i}$ are neither 
promoted to spurions nor substituted by a dynamical field (i.e. $s$ or $s^*$), and this represents a difference with respect to the 
analysis in Ref.~\cite{Brivio:2017sdm}.

Accordingly to the charge assignment in Tab.~\ref{tab:charges}, the PQ-breaking terms in the fermionic CW potential, $\hat{d}_i$, are 
vanishing, while the $SO(5)$ breaking terms read
\begin{equation}
\tilde{d}_1 = 0\,,\qquad\qquad 
\tilde{d}_2 = y_2^2 \Lambda_1^2 \,.
\end{equation}
In consequence, in this scenario, only a log-divergent $SO(5)$ breaking contribution to the $h$--mass term arises from the 
fermionic part of the CW potential, while no $\s$ tadpole contribution is generated. This is different from the analysis performed 
in Ref.~\cite{Feruglio:2016zvt}, where the only $SO(5)$ symmetry breaking terms considered have been the $\sigma$ tadpole and the 
$h^2$ terms. The minimisation of the scalar potential performed in Ref.~\cite{Feruglio:2016zvt} does not apply to this model and a 
new analysis is in order. To obtain a viable $SO(5)$ and EW spontaneous symmetry breaking at least two different $SO(5)$ breaking 
terms are necessary. Additional unavoidable sources of $SO(5)$ breaking comes from the gauge sector, as shown in Eq.~(\ref{tracemm2g}). 
The minimal counter--term potential required at tree--level by renormalisability of the theory, once the charge assignment has been 
chosen, is then given in the unitary gauge by
\be
V^\text{c.t.}(h,\sigma)=-\beta f^2 h^2 + \gamma h^4\,.
\label{EffectiveCWV}
\ee

Other values for the PQ charges are possible by changing the explicit value of $n_s$, but they lead to the same physical model presented 
above, at least for what concerns the $SO(5)/SO(4)$ phenomenology and the analysis of the scalar potential. The physical dependence 
on the explicit value of $n_s$, and then of those of the exotic fermions, can be found in the couplings between the axion and the gauge 
field strengths, whose coefficients are determined by the chiral anomaly (see Refs.~\cite{Dimopoulos:1979es,Giudice:2012zp,Redi:2012ad,
Redi:2016esr,DiLuzio:2016sbl,Farina:2016tgd,Ema:2016ops,Calibbi:2016hwq,DiLuzio:2017pfr,Coy:2017yex,Arias-Aragon:2017eww} for other 
studies where the axion couplings are modified with respect to those in the original KSVZ model). 

The explicit expression describing the Lagrangian modification under generic PQ transformations are reported in the 
App.~\ref{App.GenericPQTransf}. The coefficients of the axion couplings with the gauge boson field strengths in the physical basis,
\begin{equation}
\begin{split}
\delta\LL\supset&-\dfrac{\alpha_s}{8\pi}\,\frac{a}{f_a} c_{agg} G^a_{\mu\nu}\tilde{G}^{a\mu\nu}-\dfrac{\alpha_{em}}{8\pi}\,
\frac{a}{f_a}c_{a\gamma\gamma}F_{\mu\nu}\tilde{F}^{\mu\nu}-\dfrac{\alpha_{em}}{8\pi}\,\frac{a}{f_a}c_{aZZ}Z_{\mu\nu}\tilde{Z}^{\mu\nu}+\\
&-\dfrac{\alpha_{em}}{8\pi}\,\frac{a}{f_a}c_{a\gamma Z}F_{\mu\nu}\tilde{Z}^{\mu\nu}-\dfrac{\alpha_{em}}{8\pi}\,
\frac{a}{f_a}c_{aWW}W^+_{\mu\nu}\tilde{W}^{-\mu\nu}\,,
\end{split}
\label{PhysicalAxionCouplings}
\end{equation}
are reported in Tab.~\ref{tab:pheno} for the PQ scenario under consideration\footnote{In the present discussion, only one fermion 
generation has been considered. Once extending this study to the realistic case of three generations~\cite{USFUTURE}, the values 
reported in Tab.~\ref{tab:pheno} will be modified: for example, assuming that the same charges will be adopted for all the 
fermion generations, the numerical values in the table will be multiplied by a factor 3.}. It will be useful for the future 
discussion to introduce the notation of the effective couplings
\be
g_{agg}\equiv \dfrac{\alpha_{s}}{2\pi}\,\frac{c_{agg}}{f_a}\qquad\qquad
g_{i}\equiv \dfrac{\alpha_\text{em}}{2\pi}\,\frac{c_{i}}{f_a}\,,
\label{gAxionEffectiveCouplings}
\ee
where $i=\{a\gamma\gamma,\,aZZ,\,a\gamma Z,\,aWW\}$.

\begin{table}[h!]
\begin{center}
\begin{tabular}{|c|c|c|c|c|}
\hline
$c_{agg}$ & $c_{a\gamma\gamma}$ & $c_{aZZ}$ & $c_{a\gamma Z}$ &  $c_{a W W}$\\
\hline
 $8$ & $112/3$ & $49.3$ & $ 17.8$ & $108.1$\\
\hline
\end{tabular}
\caption{\em The coefficients of the axion couplings to the gauge boson field strengths in the physical basis are reported, where the normalisation is defined in Eq.~(\ref{PhysicalAxionCouplings}).}
\label{tab:pheno}
\end{center}
\end{table}

The charge assignment in Tab.~\ref{tab:charges} corresponds to the minimal setup among all the possible AML$\s$M constructions, where 
the minimality refers to the number of new parameters that are introduced with respect to the ML$\s$M: the number of parameters in the 
fermionic Lagrangian is the same; in the scalar potential, only three additional parameters are considered, corresponding to the 
PQ sector ($f_s$, $\lambda_s$ and $\lsp$), and in particular only two $SO(5)$ breaking terms are present (corresponding to $\beta$ 
and $\gamma$); the PQ charges also represent degrees of freedom and the minimal model in Tab.~\ref{tab:charges} is univocally determined 
by fixing $n_s$. Indeed, conditions 1 and 2 impose that the difference between the charges of the LH and RH components of the SM fermions 
is vanishing, $n_{q_L}-n_{t_R}=0$, and in consequence it is always possible to redefine the whole set of PQ charges such that 
$n_{q_L}=n_{t_R}=0$. 

It is worth mentioning that an alternative charge assignment can be found satisfying to the conditions 1-4, but this scenario is not 
minimal in terms of number of parameters. In this case, the charges are such that $n_{t_R}=n_{\chi_L}=n_{\chi_R}=n_{\psi_L}=n_{\psi_R}\mp 
n_s=n_{q_L} \mp n_s$, where the ``$-$'' or ``$+$'' refer to the presence of $z_5$ or $\tilde{z}_5$ terms in the Lagrangian, respectively. 
As discussed in Ref.~\cite{Brivio:2017sdm}, SM fermions transform under the PQ symmetry, differently from the minimal AML$\s$M in 
Tab.~\ref{tab:charges}. Moreover, the Dirac mass term $M_1$ is allowed in the Lagrangian, while the $\psi$ fermions receive mass from 
the Yukawa-like term proportional to $z_5$ (or $\tilde{z}_5$). Moreover, the terms proportional to $\Lambda_{1,2,3}$ and $y_1$ are 
allowed, while the one with $y_2$ is forbidden. In consequence, the term $\tilde{d}_1$ in Eq.~(\ref{fbreaking}) is not vanishing and 
then a $\sigma$ tadpole needs to be also added into the counter term potential $V^\text{c.t.}(h,\s)$. The number of $SO(5)$ breaking 
parameters is now increased by one unit with respect to the minimal case discussed above. For this reason, this second scenario is not 
considered in what follows.

\section{The Scalar Potential}
\label{Sect.ScalarPotential}

As constructed in the previous section, the tree--level renormalisable scalar potential of the minimal AML$\s$M reads
\begin{equation}
V(h,\sigma, r) = \lambda (h^2 + \sigma^2 -f^2)^2 -\beta f^2 h^2 + \gamma h^4 + \lambda_s (r^2 - f_s^2)^2 - \lsp\,r^2\,(h^2 + \sigma^2)\,.
\label{renpot}
\end{equation}
When $f^2>0$ and $f_s^2>0$, the minimum of the potential allows for the $SO(5)$, $\UPQ$ and EW spontaneous symmetry breaking 
with non-vanishing VEVs, 
\be
\begin{aligned}
v^2_h=&  \dfrac{\beta}{2 \gamma} f^2 \\
v^2_\s=&\left(1- \frac{\lsp^2}{4\lam\lam_s}\right)^{-1} \left\{f^2 \left[ \left(1-\frac{\beta}{2\gamma}\right) + 
           \frac{\beta}{2\gamma}\frac{\lsp^2}{4 \lam\lam_s}\right] + 
           \frac{f_s^2}{2} \dfrac{\lsp}{\lam} \right\}  \\
v^2_r=&\left(1- \frac{\lsp^2}{4\lam\lam_s}\right)^{-1} \left\{ f_s^2 + \frac{f^2}{2} \frac{\lsp}{\lam_s} \right\}\equiv f^2_a \,,
\end{aligned}
\label{GenericVEVs}
\ee
where the condition $v_r \equiv f_a$ is imposed to have canonically normalised axion kinetic term, see Eqs.~(\ref{Lagr}) and 
(\ref{favsvr}). For sake of definiteness we will indicate in the following with $\hat{h}$, $\hat{\s}$ and $\hat{r}$ the physical 
fields after SSB breaking. Assuming all parameters non--vanishing, the following conditions on the parameters must be imposed:  
\begin{enumerate}[(i)]
\item $\lambda>0$ and $\lambda_s>0$ in order to lead to a potential bounded from below.
\item $\beta$ and $\gamma$ should have the same sign in order to guarantee a positive $v_h^2$ value. Following the sign 
convention adopted in Eq.~(\ref{renpot}), when both parameters are positive, the explicit symmetry breaking terms sum 
``constructively'' to the quadratic and quartic terms in the potential in the broken phase, and a larger parameter space 
is allowed. Moreover, the ratio $\beta/2\gamma<1$ leads to $v_h<f$, corresponding to the expected ordering in the symmetry breaking scales. 

\item $\lsp$ should satisfy to
\be
\lsp^2 < 4 \lam \lam_s
\label{lspcondition}
\ee 
in order to enforce positive $v_\s^2$ and $v_r^2$ values. For negativa $\lsp$ values, additional constraints could be enforced 
depending on the values of the other parameters. The sign convention chosen in Eq.~(\ref{renpot}) guarantees that no cancellation 
in $v_\s^2$ and $v_r^2$ occurs for $\lsp>0$.
\end{enumerate}

Once the symmetries are spontaneously broken, mass eigenvalues and eigenstates can be identified. While the general case can 
be studied only numerically (see Sect.~\ref{Sect.Pheno}), simple analytical expressions can be obtained in two specific frameworks:
\begin{enumerate}
 \item Integrating out the heaviest scalar dof, whose largest component is the radial scalar field $r$, and 
       studying the LO terms of the Lagrangian;
 \item Assuming $f_s\sim f$, expanding perturbatively in the small $\beta$ and $\lsp$ parameters.
\end{enumerate}
These two cases will be discussed in the following sections.

\subsection{Integrating Out The Heaviest Scalar Field}

A clear hierarchy between the three mass scalar eigenstates is achievable for large values of $\lambda_s$ and/or $f_s$: the mass 
of the heaviest scalar dof receives a LO contribution proportional to
\be
m_3\propto \sqrt{8\lam_s} f_s\,.
\label{MassHeaviest}
\ee 
With increasing values of $\lambda_s$ and/or $f_s$, the contaminations of $\hat{h}$ and $\hat{\s}$ to the heaviest scalar dof, 
in this region of the parameter space, tend to vanish and the only relevant constituent is the radial component, $\hat r$. From  
the expression in Eq.~(\ref{MassHeaviest}), one can envisage two different ways for integrating out the heaviest dof, either taking 
the limit $\lam_s \gg 1$ or taking $f_s \gg f \sim \sqrt{s_\text{cm}}$, being $\sqrt{s_\text{cm}}$ the typical centre of mass energy scale 
at LHC. These two cases represent two physically different scenarios that are discussed separately. 

The case for $\lam_s\gg1$, with $f_s$ of the same order of $f$, corresponds to the $\UPQ$ non-linear spontaneous symmetry 
breaking framework\footnote{In the case where an UV strong interacting dynamics is responsible of the largeness of $\lam_s$, 
new resonances are expected at the scale $\lesssim 4\pi f_s$ (see the naive dimensional analysis \cite{Manohar:1983md}).}: this is the 
traditional axion frameworkwhere the only component of $s$ in the low-energy spectrum is the axion, while $\hat{r}$ is integrated out. 
As the Yukawa-like couplings of the exotic fermions do not depend on $\lambda_s$, the decoupling of $\hat{r}$ does not have any impact 
on the spectrum of the exotic fermions, that depends exclusively of the specific value chosen for $f_s$. One can then consider in detail 
the two limiting cases: $f_s \sim f$ or $f_s \gg f$. Notice that in the second scenario, when $f_s$ is much larger than any other 
mass scale, the exotic fermion sector decouples at the same time as the heavier scalar dof. 

Considering the scalar sector, integrating out the $\hat{r}$ component, leads to an effective scalar potential that,
at LO in the appropriate expansion parameter, reads
\begin{equation}
V^{LO}_R (h,\sigma) = \lambda_R (h^2 + \sigma^2 - f_R^2)^2  - \beta_R f_R^2 h^2 + \gamma h^4\, ,
\label{VLORenorm}
\end{equation}
in terms of conveniently renormalised couplings:
\be
\lambda_R= k_\lambda \lambda\,,\qquad\qquad
\beta_R=\dfrac{k_\lambda}{k_f} \beta\,,\qquad\qquad
f_R^2=\dfrac{k_f}{k_\lambda} f^2\,.
\label{renparameters}
\ee
The finite renormalisation constants $k_\lam$ and $k_f$ are going to be different in the two limiting cases as it will be detailed 
in the following subsections. 

The minimum of the effective scalar potential in Eq.~(\ref{VLORenorm}) corresponds to the following VEVs for the lighter dofs 
$\hat{h}$ and $\hat{\s}$:
\be
v^2_h = \dfrac{\beta_R}{2\gamma}f_R^2\,,\qquad \qquad
v^2_\s = f^2_R\left(1-\dfrac{\beta_R}{2\gamma}\right)\,, 
\label{vevR}
\ee
satisfying to 
\be
v^2_h+v^2_\s=f^2_R\,.
\ee
The restrictions on the parameters that follow from Eq.~(\ref{GenericVEVs}) hold for the expressions just obtained: $\beta_R/\gamma$ 
needs to be positive in order to guarantee $v^2_h>0$; $f_R$ is required to be larger than $v_h$ to ensure $v^2_\s>0$. Moreover, 
if $v_\s>v_h$ then the field $\hat{h}$ is the largest component of the mass eigenstate that can be interpreted as the physical 
Higgs particle originated as a GB of the $SO(5)/SO(4)$ SSB mechanism.

From Eq.~(\ref{VLORenorm}) and using the relations of Eq.~({\ref{vevR}}) one derives the following mass matrix:
\be
\M^2_s =  8\, \lam_R 
\begin{pmatrix}
 (1+ \gamma / \lam_R) v_h^2 &  v_h v_\s \\
  v_h v_\s                  &  v_\s^2 
\end{pmatrix}\,,
\label{MassR}
\ee
whose diagonalisation is obtained by performing an $SO(2)$ rotation,
\be
\diag\left(m^2_1,\,m^2_2\right)=U(\vartheta)^T\M^2_s U(\vartheta)\qquad\text{with}\qquad 
U(\vartheta)=
\begin{pmatrix}
\cq & \sq \\
-\sq & \cq
\end{pmatrix}\,.
\ee
The expressions for the masses and the mixing obtained from the LO potential of Eq.~(\ref{VLORenorm}) are given by
\bea
m^2_{1,2} \!\!\! &=&\!\!\! 4\lam_R \left[ \left(1+\dfrac{\gamma}{\lam_R}\right) v_h^2 + v_\s^2  \pm 
    \sqrt{\left(1+\dfrac{\gamma}{\lam_R}\right)^2 v^4_h + 2 \left(1-\dfrac{\gamma}{\lam_R}\right) v^2_h v^2_\s + v^4_\s} \right]
\label{eigenvaluesR}\\
\tan 2\vartheta \!\!\! &=&\!\!\! \frac{2 v_h v_\s}{v_\s^2-(1+ \gamma/\lam_R) v_h^2 } \,. 
\label{mixingR}
\eea
The positivity of the two mass square eigenvalues is guaranteed imposing that both the trace and the determinant of the 
mass matrix in Eq.~(\ref{MassR}) are positive: this leads to
\be
\lambda_R>0\,,\qquad\qquad
\gamma>0\,,\qquad\qquad
\beta_R>0\,,
\ee
where the last condition follows from the requirement that $\gamma$ and $\beta_R$ should have the same sign in order to guarantee 
a positively defined $v_h^2$ value, as discussed below Eq.~(\ref{GenericVEVs}).

The following two subsections will describe in detail the two limits $\lambda_s\gg 1$ and $f_s\gg f \sim \sqrt{s_\text{cm}}$, focusing on 
the scalar sector.

\boldmath
\subsubsection*{The Large PQ Quartic Coupling: $\lam_s \gg1$ and $f_s \sim f$}
\unboldmath

For $\lam_s$ in the strongly interacting regime, the radial component $r$ can be expanded in inverse powers of $\lam_s$ (see 
Ref.~\cite{Gavela:2016vte} for a similar analysis): at the NLO, one has
\be
r = f_s +\dfrac{1}{\lam_s}r_1\,.
\ee
Solving the Equations Of Motion (EOMs) perturbatively allows to determine $r_1$:
\be
r_1 = \frac{\lsp}{4f_s}\left(h^2+\s^2\right)  +\frac{1}{8f^3_s}\left(\dmu a\right) \left(\dmup a\right)\,.
\ee
The effective Lagrangian at the NLO reads
\be
\begin{split}
\LL_s  =&  \frac{1}{2}(\dmu h)(\dmup h) + \frac{1}{2}(\dmu \s)(\dmup \s) - \frac{h^2}{4} \Tr\left( \V^\mu \V_\mu\right) +\\
&+\frac{1}{2}(\dmu a)(\dmup a)- \lambda_R \left(h^2+\s^2 -f_R^2\right)^2 + \beta_R f_R^2 h^2 - \gamma h^4 + \delta\LL^\text{NLO}_s
\end{split}
\label{EffectiveLagrIntegratedOut}
\ee
with $\lambda_R$, $\beta_R$ and $f_R^2$ defined as in Eq.~(\ref{renparameters}) with 
\be
k_\lam =1\,,\qquad\qquad
k_f = \left( 1 + \frac{1}{2} \frac{\lsp}{\lam} \frac{f^2_s}{f^2} \right)\,,
\label{fRNL}
\ee
and where the NLO correcting term is given by
\be
\delta\LL_s^\text{NLO} = \dfrac{4}{\lam_s} f_s^2 r_1^2= \frac{\lsp^2}{4\lam_s} \left[(h^2+\s^2) + 
                         \frac{1}{2 f_s^2} \left(\dmu a\right) \left(\dmup a\right) \right]^2\,.
\label{EffectiveLagrIntegratedOutDelta}
\ee
In this scenario, $f_R$ is the new effective $SO(5)/SO(4)$ breaking scale, while the $SO(5)$ quartic coupling $\lambda=\lambda_R$ 
remains unchanged. The positivity of $f_R^2$ translates into a constraint on the couplings $\lsp$:
\be
\lsp > - 2 \lam \frac{f^2}{f^2_s}\,,
\label{lspConstraint}
\ee
where $\lam$, $f^2$ and $f^2_s$ are all positive (see the discussion at the beginning of Sect.~\ref{Sect.ScalarPotential}). The value 
$\lsp=0$ is special: $\lsp$ represents the portal between the $SO(5)$ and the PQ sectors, and therefore once it is vanishing the 
two sectors are completely decoupled.

A convenient limit that will be used to compare with the numerical analysis, is when $\lambda_s\gg\lambda_R\gtrsim1$ and small $\beta$, 
for which the expressions in Eqs.~(\ref{eigenvaluesR}) and (\ref{mixingR}), reduce to
\bea
m_1^2 &=& 4\beta f^2\left(1-\dfrac{\beta}{2\gamma}\right) \\
m_2^2 &=& 8\lambda f^2\left(1+\dfrac{\beta^2}{4\gamma\lambda}\right)+4\lsp f_s^2
\label{MassesIntegrateOutr_lambdasgglambdagg1}
\eea
with the mixing angle defined as
\bea
\tan2\vartheta = \left(1-\dfrac{\beta}{\gamma}\right)^{-1}\sqrt{\dfrac{2\beta}{\gamma}\left(1-\dfrac{\beta}{2\gamma}\right)} \,.
\eea

\boldmath
\subsubsection*{The Large PQ SSB Scale: $f_s \gg f \sim \sqrt{s_\text{cm}}$}
\unboldmath

In the limit $f_s\gg f \sim \sqrt{s_\text{cm}}$, being $\lam_s$ in either the perturbative or strongly interacting regimes, a similar 
expansion as in the previous subsection can be performed on the field $r$, adopting as new dimensionless expanding parameter $f/f_s$. 
Within this setup $r$ at NLO reads 
\be
r = f_s + \dfrac{f}{f_s}\, r_1 \,.
\ee
Solving the EOMs in this case, one gets
\be
r_1 =\dfrac{\lsp}{4\lambda_sf}\left(h^2+\sigma^2\right)\,.
\ee
Once substituting these expressions in Eq.~(\ref{Lagr}), the effective Lagrangian in Eq.~(\ref{EffectiveLagrIntegratedOut}) is obtained 
with 
\be
\begin{split}
\delta\LL^\text{NLO}_s=&
\dfrac{\lsp}{4\lambda_s}\dfrac{\left(h^2+\s^2\right)}{f_s^2}\left[\left(\partial_\mu a\right)\left(\partial^\mu a\right)+
             \dfrac{\lsp^2}{4\lambda_s}\left(h^2+\s^2\right)^2\right]+\\
            &+\frac{\lsp^2}{32\lambda_s^2f_s^2}\partial_\mu\left(h^2+\s^2\right)\partial^\mu\left(h^2+\s^2\right)\,,
\end{split}
\ee
and $\lam_R$ and $f_R^2$ defined in Eq.~(\ref{renparameters}), with $k_\lam$ and $k_f$ explicitly given by
\be
k_\lam = \left( 1-\dfrac{1}{4}\dfrac{\lsp^2}{\lambda \lambda_s}\right)\,,\qquad\qquad
k_f = \left( 1 + \frac{1}{2} \frac{\lsp}{\lam} \frac{f^2_s}{f^2} \right)\,.
\label{lambda&fRfs}
\ee
An increasing value of $f_s$ corresponds to an increasing value of $f_R$.  However, caution is necessary in the case when $\lsp$ is exactly 
vanishing, as the $SO(5)$ and PQ sectors are decoupled: in this specific case $f_R=f$ and the $SO(5)$ SSB sector is not affected by the 
integration out of the radial dof $r$. 

Differently from the previous case, here also a new renormalised quartic couplings $\lam_R \neq\lam$ is introduced. To ensure a 
potential bounded from below both $f_R^2$ and $\lam_R$ need to be positive, leading to the following constraints on $\lsp$,
\be
\lsp > - 2 \lam \frac{f^2}{f^2_s} \qquad \wedge\qquad
\lsp^2<4\lambda \lambda_s \,.
\label{LambdaRCondition}
\ee

In the limiting case under discussion, the explicit values for the two lightest mass eigenvalues and for their mixing 
in Eqs.~(\ref{eigenvaluesR}) and (\ref{mixingR}), assuming for simplicity $\lsp^2 \ll \lam \lam_s$, simplify to
\be
m^2_1 = 4\beta f^2\left(1 - \frac{\beta}{\gamma}\dfrac{\lambda}{\lsp}\dfrac{f^2}{f_s^2} \right) \,,\qquad \qquad
m^2_2 = 4 \lsp f_s^2\left(1+2\dfrac{\lambda}{\lsp}\dfrac{f^2}{f_s^2} \right)
\label{MassesIntegrateOutr_fsggf}
\ee
with the mixing angle given by
\bea
\tan 2\vartheta &= 2\sqrt{\dfrac{\beta}{\gamma}\dfrac{\lambda}{\lsp}} \dfrac{f}{f_s}\,.
\eea

\boldmath
\subsection{The Case for $f_s\sim f \sim \sqrt{s_\text{cm}}$ and $\beta,\lsp\ll 1$}
\unboldmath

For $f_s\sim f \sim \sqrt{s_\text{cm}}$, all the three scalar dofs are retained in the low energy spectrum and in general a stronger 
mixing between the three eigenstate  is expected, compared to the previous setups. Complete analytical expression for the masses and 
mixings cannot be written in particularly elegant and condensed form. Nevertheless, simple analytic results can be obtained under the 
assumption that $\beta,\,\lsp\ll 1$, which are natural conditions in the AML$\s$M. The first condition comes from the requirement 
that $v_h$ coincides with the EW scale $v$, defined by $v\equiv2M_W/g=246\GeV$, and it is much smaller than the $SO(5)$ SSB scale, 
i.e. $v_h<f$. The smallness of $\lsp$ follows, instead, from the assumption that the $SO(5)$ and PQ sectors are determined by two 
distinct dynamics and therefore the two breaking mechanisms occur independently. A large $\lsp$ would indicate, instead, a unique 
origin for the two symmetry breaking mechanisms and would signal the impossibility of disentangling the two sectors.

Expanding the expressions for the generic VEVs found in Eq.~(\ref{GenericVEVs}) for small $\beta$ and $\lsp$, it leads to the 
following simplified expressions:
\be
\begin{aligned}
v^2_h &= \dfrac{\beta}{2 \gamma} f^2 \\
v^2_\s &= \left(1-\frac{\beta}{2\gamma}\right)f^2 + \dfrac{\lsp}{\lam}\frac{f_s^2}{2} \,+\, 
          {\mathcal O}\left(\beta^2,\beta\lsp,\lsp^2\right) \\
v^2_r &= f_s^2 + \frac{\lsp}{\lam_s}\frac{f^2}{2}  \,+\, {\mathcal O}\left(\beta^2,\beta\lsp,\lsp^2\right) \,,
\end{aligned}
\label{vevexp}
\ee
where in the brackets the dependence on $\beta$ and $\lsp$ of the higher order corrections is reported. The scalar squared mass matrix 
is given by the following expression
\be
\M^2_s= 2 \left(
\begin{array}{ccc}
 4 (\gamma + \lambda) v_h^2& 4\lambda v_h v_\s                         & -2\lsp v_h v_r\\
 4 \lambda v_h v_\s          & 2 \lambda (v_h^2+3 v_\s^2-f^2)-\lsp v_r^2 & -2\lsp v_\s v_r\\
-2 \lsp v_h v_r              &-2 \lsp v_\s v_r                           & -\lsp(v_h^2+v_\s^2)+6\lambda_s v_r^2-2\lambda_s f_s^2
\end{array}\right)
\nn
\ee
that can be diagonalised performing an orthogonal transformation,
\be
\diag\left(m_1^2,\,m_2^2,\,m_3^2\right)=U(\vartheta_{12},\vartheta_{23})^T \M^2_s U(\vartheta_{12},\vartheta_{23})
\ee
with
\be
U(\vartheta_{12},\vartheta_{23}) = U(\vartheta_{12}) U(\vartheta_{23})\,,
\ee
the product of a rotation in the 12 sector and in the 23 sector respectively, of angles $\vartheta_{12}$ and $\vartheta_{23}$. 
The resulting mass eigenvalues read
\be
\begin{aligned}
m_1^2 =& 4\beta f^2\left(1-\dfrac{\beta}{2\gamma}\right) + {\mathcal O}\left(\beta^3,\beta^2\lsp\right) \\
m_2^2 =& 8\lam f^2 \left(1+\frac{1}{2}\frac{\lsp}{\lam}\frac{f_s^2}{f^2}\right) + 
              {\mathcal O}\left(\beta^2,\beta\lsp,\lsp^2\right) \\
m_3^2=& 8\lambda_sf_s^2\left(1+\frac{1}{2}\frac{\lsp}{\lam_s}\frac{f^2}{f_s^2}\right) + 
              {\mathcal O}\left(\beta\lsp,\lsp^2\right)\,,
\end{aligned}
\label{Massesfssimf}
\ee
while the mixing angles are given by
\be
\tan 2\vartheta_{12}= \sqrt{\dfrac{2\beta}{\gamma} } \left(1\,+\, {\mathcal O}(\beta, \lsp) \right) \,,\quad
\tan 2\vartheta_{23}= \dfrac{f f_s}{\lambda_s f_s^2-\lambda f^2} \, \lsp \left(1 \,+\, {\mathcal O}(\beta,\lsp)\right) \,.
\label{mixingII}
\ee
As for Eq.~(\ref{vevexp}), only the first two relevant terms in the expansion are reported in the expressions in Eqs.~(\ref{Massesfssimf}), 
while the powers in $\beta$ and $\lsp$ of the expected next order terms are shown in the brackets. Instead, in the formula for the mixing 
angles in Eq.~(\ref{mixingII}), only the first term is indicated. Notice that, once considering the next order terms in the masses 
expressions, a rotation in the 13 sector is also necessary to exactly diagonalise the squared mass matrix.

\subsection{Numerical Analysis}
\label{Sect.Pheno}

This subsection illustrates the numerical analysis on the parameter space of the scalar potential. The analytic results of the 
specific cases presented in the previous subsection will be used to discuss the numerical outcome. To this aim, a more general 
notation with respect to the one previously adopted is introduced. The scalar mass matrix $\M_s$ is real and can be diagonalised 
by an orthogonal transformation,
\begin{equation}
\diag (m_1^2,m_2^2,m_3^2) = U(\vartheta_{12},\vartheta_{23},\vartheta_{13})^T \M_s^2\, U(\vartheta_{12},\vartheta_{23},\vartheta_{13})\,,
\end{equation}
where $U(\vartheta_{12},\vartheta_{23},\vartheta_{13})\equiv U(\vartheta_{23}) U(\vartheta_{13}) U(\vartheta_{12})$ is the product 
of three rotations in the 23, 13, and 12 sectors respectively, of angles $\vartheta_{23}$, $\vartheta_{13}$ and $\vartheta_{12}$. 
The scalar mass eigenstates $\varphi_1$, $\varphi_2$, and $\varphi_3$ are defined by
\begin{equation}
	\begin{pmatrix}
	\varphi_1 \\ \varphi_2 \\ \varphi_3
	\end{pmatrix} =
	U(\vartheta_{12},\vartheta_{23},\vartheta_{13})^T 
	\begin{pmatrix}
	\hat{h} \\ \hat{\sigma} \\ \hat{r}
	\end{pmatrix}
\end{equation}
in terms of the three physical shifts around the vacua. Unless explicitly indicated, in the analysis that follows, $\varphi_1$ will be 
identified with the Higgs particle and the deviations of its couplings from the SM predicted values are interesting observables at colliders. 
The $\varphi_1$ couplings to the SM gauge bosons can be deduced from the couplings of $\hat{h}$, as $\hat\s$ and $\hat r$ are singlets 
under the SM gauge group. The composition of $\hat{h}$ in terms of $\varphi_i$ is explicitly given by
\be
\hat{h} = c_{12} c_{13} \varphi_1 + c_{13} s_{12} \varphi_2 + s_{13} \varphi_3 \equiv C_1 \varphi_1 + C_2 \varphi_2 + C_3 \varphi_3\,,
\label{hhattovarphi}
\ee
where $c_{ij}$ and $s_{ij}$ stand for $\cos\theta_{ij}$ and $\sin\theta_{ij}$, and the coefficients $C_i$ in the last equality 
have been introduced for shortness. The couplings with the SM gauge bosons can be written as
\be
\begin{aligned}
\frac{g^2}{4} (\hat{h} + v_h)^2 W^+_\mu W^{-\mu} &= m_W^2 \left( C_1 \frac{\varphi_1}{v_h}  + C_2 \frac{\varphi_2}{v_h} + 
            C_3 \frac{\varphi_3}{v_h}  +1 \right)^2 W^+_\mu W^{-\mu}\,,\\
\frac{g^2+g^{\prime2}}{8} (\hat{h} + v_h)^2 Z_\mu Z^\mu &= \dfrac{m_Z^2}{2} \left( C_1 \frac{\varphi_1}{v_h}  + 
            C_2 \frac{\varphi_2}{v_h} + C_3 \frac{\varphi_3}{v_h}  +1 \right)^2 Z_\mu Z^\mu\,.
\end{aligned}
\label{aWWaZZ}
\ee
Finally, the $\varphi_1$ couplings to the longitudinal components of $W$ and $Z$ are modified with respect to the SM predictions for the 
Higgs particle by factor of $C_1$. 

To have a clear comparison with CHM predictions, one can write the expression for the $C_1$ parameter obtained integrating out all the 
scalar dofs of our model, but the physical Higgs. The most immediate way to obtain such a result is to start from Eq.~(\ref{mixingR}) and 
expanding it for $\lambda_R\gg1$, giving 
\be
C_1\simeq1-\dfrac{1}{2}\dfrac{v_h^2}{v_\sigma^2}\equiv 1-\dfrac{\xi}{2}\,,
\label{C1expanded}
\ee
The last term on the right-hand side introduces the parameter $\xi$, that customary defines the tension between the EW and the composite scales. 
This parameter often appears in CHMs to quantify the level of non-linearity of the model. The expression in Eq.~(\ref{C1expanded}) agrees with 
previous MCHM results present in literature, see for example Ref.~\cite{Bellazzini:2014yua}. Therefore, the corresponding bounds on $\xi$, 
as the ones from Refs.~\cite{Panico:2015jxa,Khachatryan:2016vau},
\be
\xi \lesssim 0.18\quad @\,2\s \,,
\label{XiConstraint}
\ee
strictly apply to the model presented here only in the MCHM limit, i.e. when all the scalar fields, but the Higgs, are extremely massive and can be 
safely integrated out. If this is not the case, the coefficient $C_1$ is a complicate function of all the scales and parameters effectively 
present in the model.

\subsubsection*{The Scalar Potential Parameter Space}
\label{Sect.PhenoI}

The parameter space of the scalar sector is spanned by seven independent parameters: five dimensionless coefficients $\lambda$, $\lambda_s$, 
$\beta$, $\gamma$, $\lsp$, and two scales $f$ and $f_s$. By using the known experimental values of the Higgs VEV, $v_h=v\equiv246\GeV$, 
and mass $m_1=m_h\equiv125\GeV$, two of these coefficients can be eliminated in terms of the remaining five. The adopted procedure for the 
numerical analysis is to express $\gamma$ as function of $\beta$ and $f$, by inverting the $v_h^2$ expression in Eq.~(\ref{GenericVEVs}): 
\be
\gamma = \left( \frac{f}{v_h} \right)^2 \frac{\beta}{2}\,.
\ee
and then extract $\beta$, in terms of the remaining five parameters, by numerically solving the equation $m_1(\beta,\lambda,\lambda_s,
\lsp, f, f_s)=m_h$. Consequently, predictions for all the remaining observables can be obtained by choosing specific values for
$(\lambda, \lambda_s, \lsp, f, f_s)$.

\begin{figure}[t!]
\centering
\includegraphics[width=15cm]{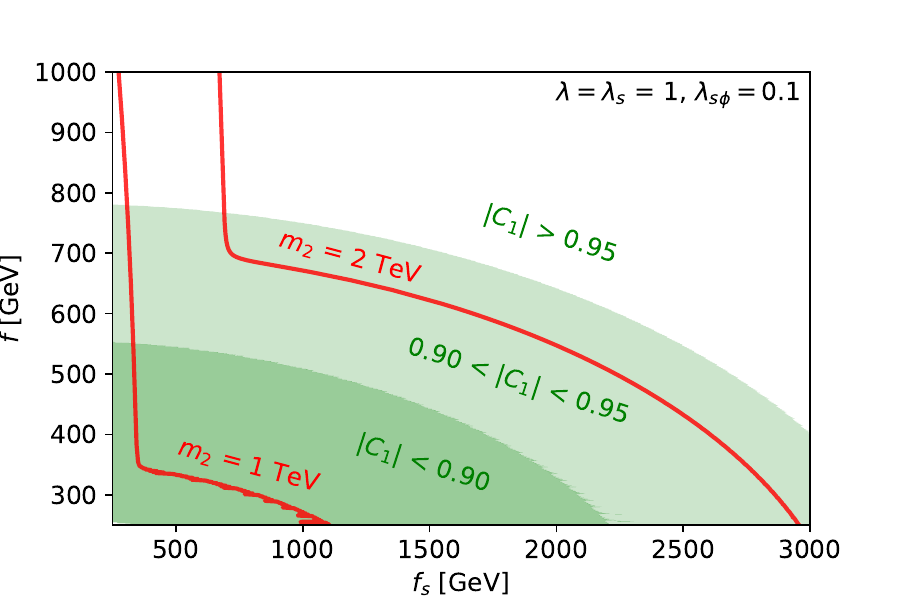}
\caption{\it \footnotesize $C_1$ contours in the $(f_s,f)$ plane, for $\lambda = \lambda_s =1$ and $\lsp = 0.1$. 
The dark green region corresponds to $|C_1| < 0.90$, while the light green one to $0.90 < |C_1| < 0.95$. In the white region $|C_1|>0.95$. The two red curves correspond to values for the next to lightest scalar $m_2 = 1$ TeV and $m_2 = 2$ TeV respectively, being the Higgs mass fixed to the reference value $m_h =125$ GeV.}  
\label{f_vs_fs}
\end{figure}

In Fig.~\ref{f_vs_fs} the bounds on the $|C_1|$ parameter in the $(f_s,f)$ plane for $\lambda=\lambda_s=1$ and $\lsp=0.1$ are shown.
The dark green region corresponds to $|C_1|<0.90$, while the light green one to $0.90<|C_1|<0.95$. In the white region $|C_1|>0.95$. 
From this plot one can have an order of magnitude comparison with present/future experimental bound on the Higgs--gauge boson interaction. 
The following bounds on $hZZ$ and $hWW$ couplings are obtained by \cite{CMSHiggs2018}, using the so called $\kappa$-framework\footnote{Notice 
that in the $\kappa$-framework one assumes that there are no new particles contributing to the $ggH$ production or $H \to \gamma \gamma$ decay 
loops.}: 
\bea
|\kappa_{Z}| &=& 0.89+0.09-0.08 \quad @ \quad 1\s \nn \\
|\kappa_{W}| &=& 1.00+0.00-0.05 \quad @ \quad 1\s 
\label{expks}
\eea
The expressions in Eq.~(\ref{aWWaZZ}) enforce the relation $\kappa_Z=\kappa_W=C_1$.

Fig.~\ref{f_vs_fs} gives the idea of the interplay between the two scales $f$ and $f_s$ for fixed values of the remaining 
adimensional parameters. For $f_s=1$ TeV, LHC can already start to exclude values of $f \lesssim 0.7$ TeV. However, for the larger value 
$f_s=3$ TeV, even values of $f\approx 0.5$ TeV will lie outside LHC exclusion reach and no precise bound separately on $f$ or $f_s$ 
can be inferred from the sole measurement of the Higgs couplings to gauge bosons, for most of the parameter space\footnote{Limits on the 
scale $f$ from EWPO will be discussed in the following section.}. Only when $\lam, \lam_s \gg 1$ are taken, the extra scalar dofs are decoupled and the CHM relation of Eq.~(\ref{C1expanded}) can be exploited. These results are compatible with the ones of Ref.~\cite{Feruglio:2016zvt}, 
where a detailed study on the allowed range for $f$ has been performed in the context of the ML$\s$M. For completeness in Fig.~\ref{f_vs_fs} also the curves corresponding to two values of the mass of the next to lightest scalar, $m_2 = 1$ TeV and $m_2=2$ TeV, have been depicted.  

In the following analysis the value $f = 2$ TeV has been chosen as benchmark. The parameter space for the remaining four variable, 
$\lambda, \lambda_s, \lsp, f_s$, can be studied, plotting the behaviour of the scalar mass eigenvalues $m_i$ and of the mixing 
coefficients squared $C_i^2$. 

\begin{figure}[h!]
\centering
\includegraphics[width=8cm]{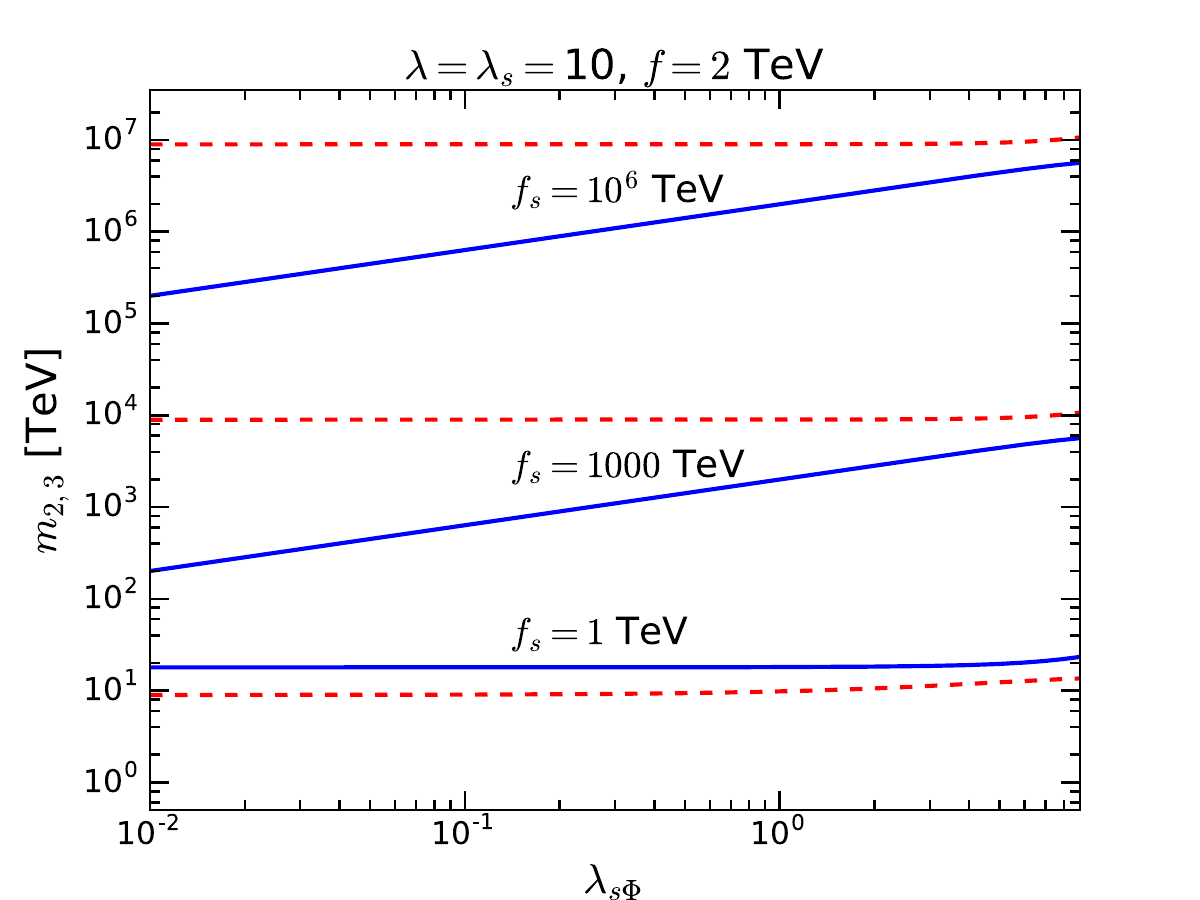}
\includegraphics[width=8cm]{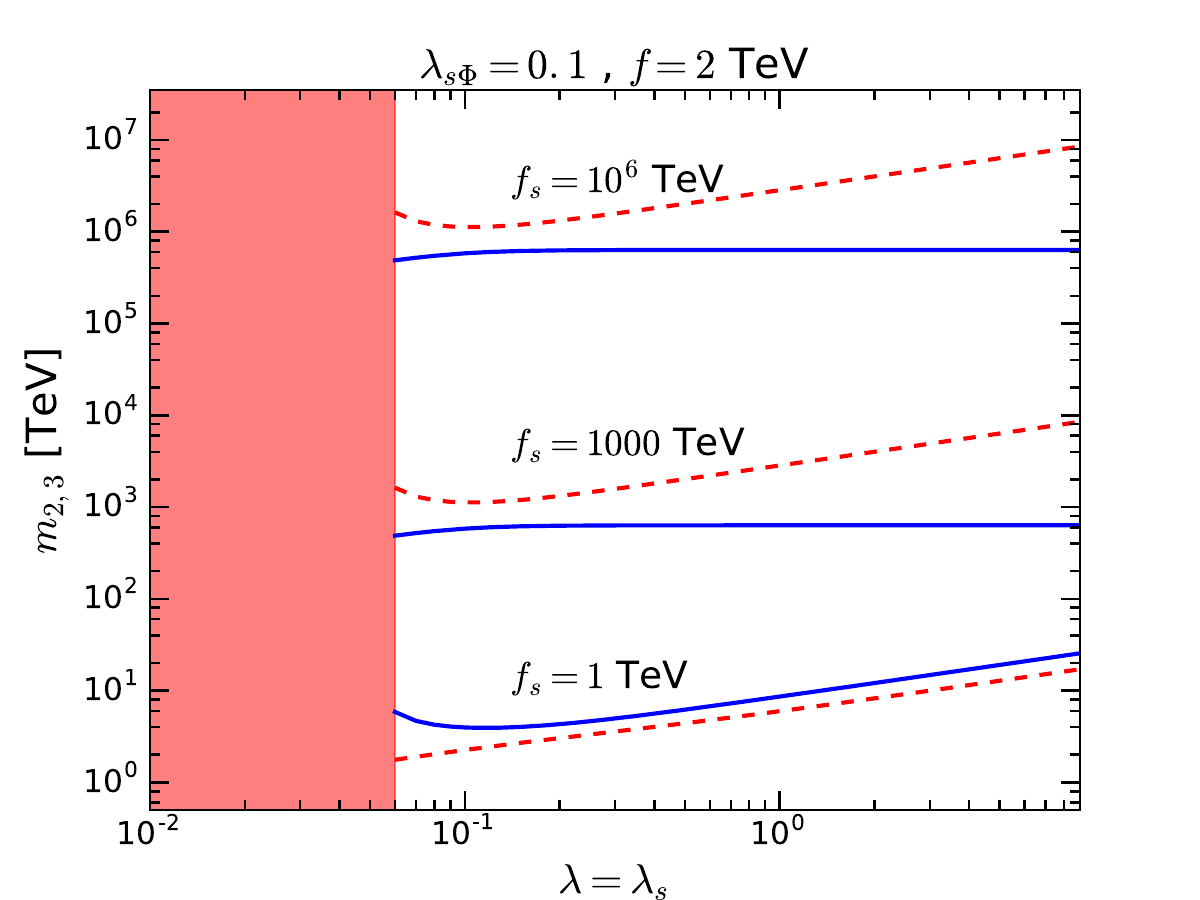}
\includegraphics[width=8cm]{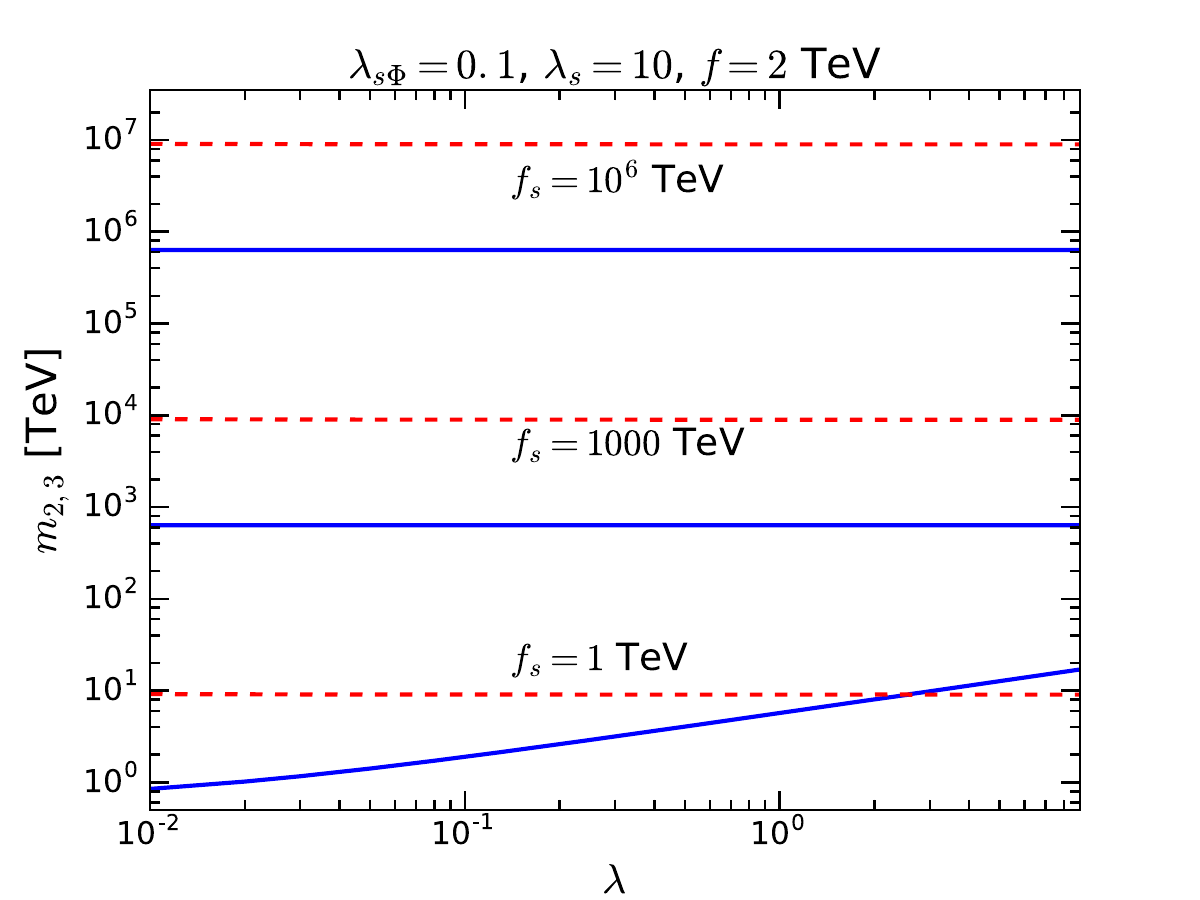}
\caption{\it \footnotesize The profiles of the scalar masses $m_2$ and $m_3$ as a function of $\lsp$ (upper left), $\lambda=\lambda_s$ 
(upper right), and $\lambda$ (lower). The other parameters are chosen at fixed values: $f = 2\TeV$; $f_s = 1\TeV,\,10^3\TeV,\,10^6 \TeV$; 
$\lambda = \lambda_s = 10$ (upper left);  $\lsp=0.1$ (upper right); $\lambda_s=10$ and $\lsp=0.1$ (lower). The red-dashed line represents 
the heaviest dof with mass $m_3$, while the blue-continue line the next-to-heaviest dof with mass $m_2$. The lightest dof is identified 
with the Higgs particle with mass $m_1=m_h$. The red area is excluded from the constraint in Eq.~(\ref{lspcondition}).}  
\label{vs_lsf}
\end{figure}

In Fig.~\ref{vs_lsf}, the masses $m_2$ and $m_3$ are shown as a function of $\lambda_{s\phi}$ (upper left), or $\lambda=\lambda_s$ 
(upper right), or $\lambda$ (lower). The mass $m_1$ is fixed at $m_h$, while the scale $f$ is taken at $2\TeV$. Three distinct values 
for $f_s$ are considered, $f_s=1\TeV,\,10^3\TeV,\,10^6\TeV$, and are shown in the same plot spanning a different region of the parameter 
space. In the plot in the upper left, the values for $\lambda$ and $\lambda_s$ are taken to be equal to $10$; in the plot in the 
upper right, $\lsp=0.1$; in the lower plot, $\lsp=0.1$ and $\lambda_s=10$. 

All these plots present features discussed in the different limiting cases of the previous section. In the three plots, the lines 
corresponding to $f_s=10^3\TeV$ and $f_s=10^6\TeV$ well represent the expressions for the masses in Eq.~(\ref{MassesIntegrateOutr_fsggf}). 
In the upper left plot, the red-dashed line represents the heaviest dof with a constant mass according with Eq.~(\ref{MassHeaviest}); 
the blue-continue line corresponds to the second heaviest dof and it shows an increasing behaviour with a constant slope, corresponding 
to the expression for $m_2^2$ that in first approximation is proportional to $\lsp$. In the upper right plot, the red area is excluded 
according to Eq.~(\ref{lspcondition}): close to this region, the analytic expressions do not closely follow the numerical results, as 
it appears in the behaviour of the red-dashed line that increases with a constant slope according to Eq.~(\ref{MassHeaviest}) only for 
$\lambda=\lambda_s\gtrsim0.1$. The blue-continue line is almost constant, as expected from the expression of  $m_2^2$ in 
Eq.~(\ref{MassesIntegrateOutr_fsggf}), except for the region with small $\lambda=\lambda_s$. In the lower plot, both the red-dashed 
and the blue-continue lines are horizontal, as expected having fixed both $\lambda_s$ and $\lsp$.

When $f_s=1\TeV$, the numerical results agree with the analytic expressions in Eqs.~(\ref{MassesIntegrateOutr_lambdasgglambdagg1}) 
and (\ref{Massesfssimf}). In the upper left plot, the red-dashed and the blue-continue lines are exchanged with respect to the lines 
for $f_s=10^3\TeV$ and $f_s=10^6\TeV$: this is in agreement with Eq.~(\ref{Massesfssimf}), as indeed for $f>f_s$ the heaviest dof 
is $\varphi_2$ and the next-to-heaviest is $\varphi_3$. Moreover, the two lines are almost horizontal as the dependence on $\lsp$ 
only enters at higher orders. In the upper right plot, both the lines increase with a constant slope, as expected from 
Eq.~(\ref{Massesfssimf}), except for small values of $\lambda=\lambda_s$, that is close to the excluded region. In the lower plot, 
the red-dashed line is almost horizontal, according to $m_3^2$ in Eq.~(\ref{Massesfssimf}), while the blue-continue line increases 
with $\lambda$, as shown by the expression for $m_2^2$. For $\lambda=2.5$ the two lines cross and $\varphi_2$ becomes the heaviest dof. 
The same conclusions are expected by analysing the expressions in Eq.~(\ref{MassesIntegrateOutr_lambdasgglambdagg1}), where $\varphi_3$ 
is integrated out: the comparison is however more difficult as $m_2^2$ depends explicitly on $\beta$ and $\gamma$, which are only 
numerically computed in terms of $\lambda, \lambda_s, \lsp, f_s$. Moreover, when $\lambda>2.5$, $\varphi_2$ should also be integrated 
out from the low-energy spectrum as its mass reaches the value of the one of $\varphi_3$, and not consistent description is expected for 
these values of $\lambda$.

\begin{figure}[h!]
\centering
\includegraphics[width=8cm]{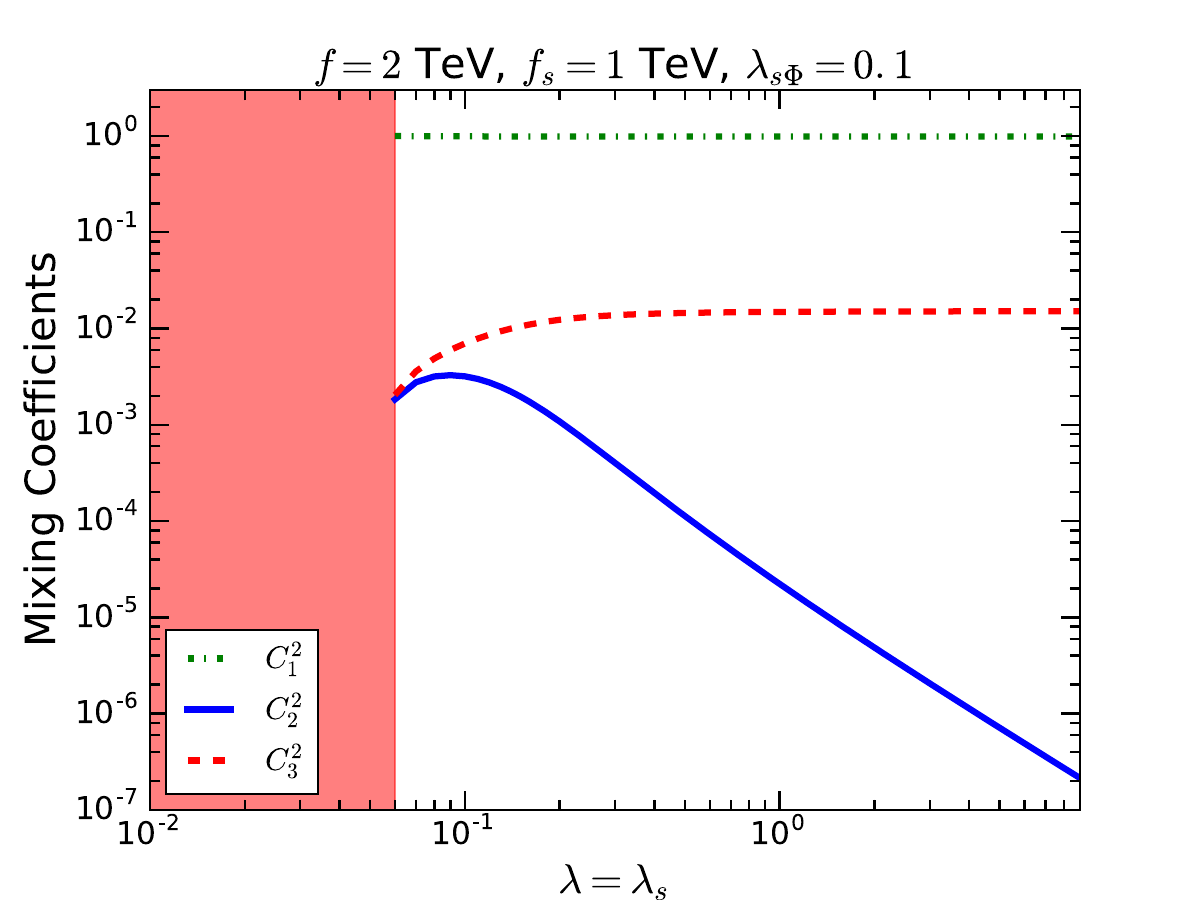}
\includegraphics[width=8cm]{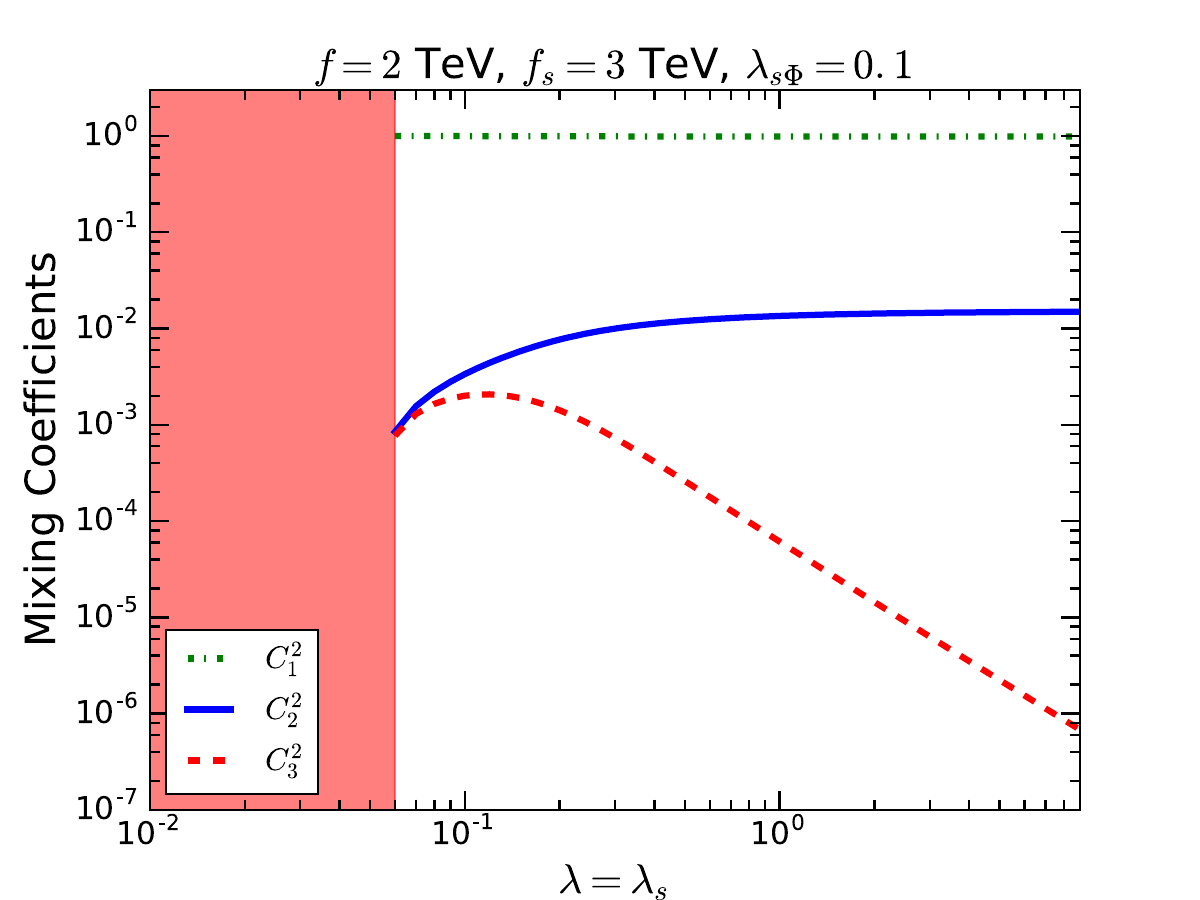}
\caption{\it \footnotesize The profiles of the coefficients squared $C_1^2$, $C_2^2$ and $C_3^2$, as a function of $\lambda=\lambda_s$. 
The other parameters are chosen at fixed values: $f = 2\TeV$; $\lsp=0.1$; $f_s = 1\TeV$ on the left and $f_s = 3\TeV$ on the right. 
The green-dot-dashed line describes $C_1^2$, the blue-continue line $C_2^2$ and the red-dashed line $C_3^2$. The red area is excluded 
from the constraint in Eq.~(\ref{lspcondition}).}  
\label{Coefficientsvs_lsf}
\end{figure}

The mixing coefficients $C_1$, $C_2$ and $C_3$ are shown in Fig.~\ref{Coefficientsvs_lsf}: the green-dot-dashed line describes $C_1^2$, 
the blue-continue line $C_2^2$ and the red-dashed line $C_3^2$. Both plots clearly show that the largest component to $\hat h$ is 
$\varphi_1$, that is identified to the physical Higgs particle. The contaminations from $\varphi_2$ and $\varphi_3$ are much smaller 
and at the level of $\sim 1\%$ at most. This is a typical feature in almost all the parameter space, and in particular for $f_s\gg f$, 
whose corresponding plots are very similar to the one in Fig.~\ref{Coefficientsvs_lsf} on the right. The only substantial difference 
between the two plots shown is the exchange behaviour between $C_2^2$ and $C_3^2$: as far as $f_s>f$ the largest contamination is 
given by $\varphi_2$, while for $f<f_s$ it is given by $\varphi_3$, as it is confirmed by Eq.~(\ref{mixingII}).

The results on the mixing coefficients can be compared to the ones for the equivalent quantities in the ML$\s$M: in the latter, only 
two scalar states are present and then only one mixing can be defined, that is between $\hat h$ and $\hat \s$; for increasing masses 
of $\varphi_2$, which almost coincides with $\hat \s$, the sibling of $C_2^2$ asymptotically approaches the ratio $v^2/f^2$ and a 
benchmark value of $0.06$ has been taken in the phenomenological analysis. From Fig.~(\ref{Coefficientsvs_lsf}), the maximal value 
that $C_2^2$ (or $C_3^2$) can take is of $0.015$: this means that some differences are expected between the two models when discussing 
the EW precision observables (EWPO) and the impact of the exotic fermions.\\

In a tiny region of the parameter space, $\varphi_2$ can be lighter than $\varphi_1$, with $m_1$ still fixed at the value $m_h$. 
This is consistent with the results in Ref.~\cite{Feruglio:2016zvt}. Although this possibility is experimentally viable, from the 
theoretical perspective it is not appealing as $m_2<m_1$ requires $\lsp\lesssim10^{-7}$, corresponding to a highly tuned situation. 
Similarly, mixing parameters larger than the typical values shown in Fig.~\ref{Coefficientsvs_lsf}, for example $C_2^2\sim0.1$, can 
only be achieved for $\lsp\lesssim 10^{-4}$, another tuned region of the parameter space. Another possibility for relatively large 
mixing parameters is for $f\sim100\GeV$ and $f_s\lesssim1\TeV$, that is very unlikely as it would correspond to the case with the 
EWSB occurring before the $SO(5)/SO(4)$ symmetry breaking. In consequence, only the case with $\varphi_2$ heavier than  $\varphi_1$ 
and values of $\lsp\gtrsim0.01$ will be considered in the following.

\section{Collider Phenomenology and Exotic Fermions}
\label{Sect.PhenoII}

Within a specific CH model setup, defined by a coset, the Higgs couplings to fermions depend on the kind of exotic fermions that enrich 
the spectrum and the chosen symmetry representations. A recent review on the $SO(5)/SO(4)$ context has been presented in 
Ref.~\cite{Panico:2015jxa} and the impact at colliders of different realisations has been analysed in Ref.~\cite{Carena:2014ria}. 
The ML$\s$M, and therefore also the AML$\s$M, seems an interpolation between the so-called $MCHM_4$ and $MCHM_5$ scenarios considered 
in Ref.~\cite{Carena:2014ria}, once only the physical Higgs is retained in the low-energy theory. Typical observables of interest at 
colliders are the EWPO, the $Zb\bar{b}$ coupling, couplings of the scalar dofs to gluons and photons~\cite{Barbieri:2007bh,Gertov:2015xma}, 
and the interactions with fermions. As they have been studied for the ML$\s$M in Refs.~\cite{Feruglio:2016zvt,Gavela:2016vte}, the aim 
of this section is to extend those results to the AML$\s$M.
\\

\noindent{\bf EWPO}\\
\indent Deviations to the SM predictions for the $T$ and $S$ parameters~\cite{Peskin:1991sw} (or equivalently $\epsilon_1$ and 
$\epsilon_3$~\cite{Altarelli:1990zd}) are expected to be relevant. In the ML$\s$M, the mixing between $\hat h$ and $\hat \sigma$ can reach 
relatively large values, $\sim 0.1$, and relevant scalar contributions to $T$ and $S$ are indeed expected. However, these contributions 
can always be compensated, in some allowed region of the parameters space, once exotic fermion contributions are included.

In the AML$\s$M, for the benchmark values chosen in the previous section, the values of the scalar sector mixing parameters result very small, 
see Fig.~\ref{Coefficientsvs_lsf}, and then the contributions to $T$ and $S$ are expected to be much less relevant. For smaller values of $f$ 
consistent with Fig.~\ref{f_vs_fs}, the $\hat h$-$\hat \sigma$ mixing slightly increases, and then larger contributions to $T$ and $S$ are 
expected. In addition, relevant contributions to the EWPO from the fermionic sector can also be present. However, exactly as happens in the 
ML$\s$M case, it is always possible to evade the $T$ and $S$ bounds in a non negligible part of the full (fermionic + bosonic) parameter space.
\\

\boldmath
\noindent{\bf $Zb\bar b$ coupling}
\unboldmath\\
\indent The modification of the $Z$ couplings to $b\bar b$ is a very good observable to test a model. The most relevant contributions 
arise from the top-partner fermion, while the ones from the heavier scalar dofs turn out to be negligible. The top-partner induces 
deviations from the SM prediction of this coupling only at the one-loop level, and the effect of these contributions is soften with 
respect to those to the EWPO previously discussed. This result holds for both the ML$\s$M and the AML$\s$M. As illustrated in 
Ref.~\cite{Feruglio:2016zvt}, it is easy to accommodate the experimental measure of the $Zb\bar b$ coupling in a large part of the 
parameter space, and therefore no relevant constraint can be deduced from this observable.
\\

\boldmath
\noindent{\bf Couplings with gauge bosons and $\s$ production at colliders}
\unboldmath\\
\indent As in the SM, no tree level $\hat hgg$ and $\hat h\gamma\gamma$ couplings are present in the AML$\s$M. However, effective interactions 
with gluons and with photons may arise at the one-loop level. In consequence, all the three scalar mass eigenstates, $\varphi_{1,2,3}$, do 
couple with gluons and photons, with their interactions weighted by the corresponding mixing coefficients $C_i^2$, according to 
Eq.~(\ref{hhattovarphi}).

As worked out in details in Ref.~\cite{Feruglio:2016zvt}, the Higgs coupling with two gluons, $\varphi_1gg$, is mainly due to the top 
contribution, as the bottom one is negligible and the exotic fermion ones tend to cancel out (due to their vector-like nature). On the other 
hand, the $\varphi_2 gg$ and $\varphi_3 gg$ couplings are suppressed by $C_2^2$ and $C_3^2$ respectively, and therefore are typically at 
least $10^{-2}$ smaller than $\varphi_1gg$. Moreover, as the top quark is lighter than $\varphi_2$ and $\varphi_3$, its contribution to 
their couplings are also suppressed, and the dominant terms arise from the exotic fermion sector.

The couplings to photons receive relevant contributions, not only from loops of top quark and of exotic fermions, but also from loops of 
massive gauge bosons. The latter are the dominant ones in the case of the physical Higgs particle, i.e. for $\varphi_1\gamma\gamma$, while 
they are suppressed by $C_2^2$ and $C_3^2$ for the heavier scalar dofs and the most relevant contributions to $\varphi_2\gamma\gamma$ and 
$\varphi_3\gamma\gamma$ are those from the exotic fermions.

These results impact on the production mechanisms of the heavier dofs at collider, that are gluon fusion or vector boson fusion. From 
Fig.~\ref{vs_lsf}, the masses for $\varphi_2$ and $\varphi_3$ are typically larger than the TeV scale, within the whole range of values 
for $f$ and $f_s$ shown in Fig.~\ref{f_vs_fs}. The lowest mass values are then potentially testable at colliders, although it strongly depends 
on the couplings with gluons and the massive gauge bosons. Ref.~\cite{Feruglio:2016zvt} concluded that, in the presence of only two scalar dofs, 
the heaviest one would be constrained only for masses lower than $0.6\TeV$ and mixing coefficient $C_2^2>0.1$. Extending this result to the 
three scalar dofs described in the AML$\s$M and considering the results presented in Fig.~\ref{vs_lsf}, the present LHC data and the future 
prospects (LHC run-2 with total luminosity of $3ab^{-1}$) are not able to put any relevant bound, or in other words the heavier scalar dofs 
have production cross sections too small to lead to any signal in the present and future run of LHC.
\\

\boldmath
\noindent{\bf Impact of the exotic fermions}
\unboldmath\\
\indent The exotic fermion masses partially depend on a distinct set of parameters with respect to those entering the scalar potential. 
While this is particularly true for the ML$\s$M, where two arbitrary mass parameters $M_{1,5}^{(\prime)}$ are introduced in the Lagrangian, 
in the minimal AML$\s$M the exotic fermion masses are controlled by $f_s$, through the parameters $z_{1,5}^{(\prime)}$ (and/or 
$\tilde{z}_{1,5}^{(\prime)}$).  The largeness of $f_s$ corresponds to large masses for these exotic fermions, consistent with the 
fermion partial compositeness mechanism. Direct detections would be probably very unlikely, while their effect would manifest in 
deviations from the SM predictions of SM field couplings. In Ref.~\cite{Feruglio:2016zvt}, the exotic fermions have been integrated 
out and the induced low-energy operators have been identified. The mayor expected effects consist in decorrelations between observables 
that are instead correlated in the SM, and the appearance of anomalous couplings: these effects are very much typical of the HEFT setup, 
where the EWSB is non-linearly realised and the Higgs originates as a GB. For an overview of these analyses see 
Refs.~\cite{Brivio:2013pma,Brivio:2014pfa,Gavela:2014vra,Brivio:2015kia,Brivio:2016fzo,Merlo:2016prs,Hernandez-Leon:2017kea}.\\

Besides the effects discussed above, it is worth to mention the possibility to investigate the Higgs nature through the physics of the 
longitudinal components of the SM massive gauge bosons. As the ML$\s$M and AML$\s$M deal with the same symmetry of the SM, no additional 
effects are expected with respect to the analyses carried out in Refs.~\cite{Espriu:2012ih,Espriu:2013fia,Delgado:2013hxa,Delgado:2014jda, 
Delgado:2017cls}.

\section{The Axion and ALP Phenomenology}
\label{Sect.AxionPhen}

The axion couplings to SM gauge bosons and fermions have been bounded from several observables~\cite{Choi:1986zw,DePanfilis:1987dk,
Bjorken:1988as,Carena:1988kr,Wuensch:1989sa,Hagmann:1990tj,Asztalos:2003px,Raffelt:2006cw,Adler:2008zza,Asztalos:2009yp,Bellini:2012kz,
Friedland:2012hj,Lees:2013kla,Armengaud:2013rta,Carosi:2013rla,Salvio:2013iaa,Clarke:2013aya,Viaux:2013lha,Aprile:2014eoa,Ayala:2014pea,
Khachatryan:2014rra,Mimasu:2014nea,Dolan:2014ska,Millea:2015qra,Vinyoles:2015aba,Aad:2015zva,Krnjaic:2015mbs,Marciano:2016yhf,
Izaguirre:2016dfi,Brivio:2017ije,Bauer:2017nlg,Anastassopoulos:2017ftl,Dolan:2017osp,Merlo:2019anv}. Two recent summaries can be found in 
Refs.~\cite{Jaeckel:2015jla,Bauer:2017ris}. In the following, only the couplings with bosons will be taken into consideration, as in 
the minimal AML$\s$M described here no direct interaction is present with SM fermions\footnote{Indirect couplings arise from the same 
mechanism that generate SM fermion masses. However, experimental constraints are present on axion couplings with only light SM fermions, 
the strongest being on axion couplings with two electrons. As in the minimal AML$\s$M only the third generation fermions are considered, 
no relevant bound can be deduced considering these constraints. This analysis is postponed to further investigation~\cite{USFUTURE}.}. 
The axion couplings strongly depend on its mass, that moreover determines whether the axion is expected to decay or not inside  
the collider. On the other side, for the ALP, mass and couplings are not related. 

The following constraints hold for both a QCD axion and an ALP. \\

\noindent{\bf Coupling to photons}\\
\indent The axion coupling to photons is bounded from both astrophysical and low-energy terrestrial data, and they depend on the axion 
mass. The most recent summary on these constraints can be found in Refs.~\cite{Jaeckel:2015jla,Bauer:2017ris}, while the last update for 
masses below tens of meV is given in Ref.~\cite{Anastassopoulos:2017ftl}: the upper bounds can be summarised as
\be
\begin{aligned}
|g_{a\gamma\gamma}|\lesssim&\,7\times 10^{-11}\GeV^{-1}\qquad
&&\text{for} \qquad m_a\lesssim10 \meV
\\
|g_{a\gamma\gamma}|\lesssim&\, 10^{-10}\GeV^{-1}\qquad
&&\text{for}\qquad 10 \meV\lesssim m_a\lesssim10\eV
\\
|g_{a\gamma\gamma}|\ll&\, 10^{-12}\GeV^{-1}\qquad
&&\text{for}\qquad 10 \eV\lesssim m_a\lesssim0.1\GeV
\\
|g_{a\gamma\gamma}|\lesssim&\, 10^{-3}\GeV^{-1}\qquad
&&\text{for}\qquad 0.1 \GeV\lesssim m_a\lesssim1\TeV\,.
\end{aligned}
\label{Constraintsa2gamma}
\ee
For masses between $10\eV$ and $0.1\GeV$, and in particular for the so-called MeV window, the coupling $g_{a\gamma\gamma}$ is constrained by (model dependent) cosmological data~\cite{Millea:2015qra}. These bounds can be translated in terms of $f_a/|c_{a\gamma\gamma}|$ 
through Eq.~(\ref{gAxionEffectiveCouplings}): taking $\alpha_\text{em}=1/137.036$,
\be
\begin{aligned}
\dfrac{f_a}{|c_{a\gamma\gamma}|}\gtrsim&\,2\times 10^{7}\GeV\qquad
&&\text{for}\qquad m_a\,\,\lesssim10 \meV
\\
\dfrac{f_a}{|c_{a\gamma\gamma}|}\gtrsim&\,10^{7}\GeV\qquad
&&\text{for}\qquad 10 \meV\lesssim m_a\,\,\lesssim10\eV
\\
\dfrac{f_a}{|c_{a\gamma\gamma}|}\gg&\,10^{9}\GeV\qquad
&&\text{for}\qquad 10 \eV\lesssim m_a\,\,\lesssim0.1\GeV
\\
\dfrac{f_a}{|c_{a\gamma\gamma}|}\gtrsim&\, 1\GeV\qquad
&&\text{for}\qquad 0.1 \GeV\lesssim m_a\,\,\lesssim1\TeV\,.
\end{aligned}
\label{Constraintsfaa2gamma}
\ee
In Ref.~\cite{Brivio:2017sdm} a dedicated analysis of the axion coupling to photons within the AML$\s$M is presented, including constraints and prospects from current experiments. Moreover, for masses $0.1\GeV\lesssim m_a\lesssim5\GeV$, the bound may be improved by two orders of magnitude with dedicated analyses on BaBar data and at Belle-II~\cite{Mimasu:2014nea,Izaguirre:2016dfi,Dolan:2017osp}. \\
    
\noindent{\bf Coupling to gluons}\\
\indent The axion coupling to gluons has been constrained by axion-pion mixing effects~\cite{Choi:1986zw,Carena:1988kr} and mono-jet 
searches at colliders~\cite{Khachatryan:2014rra,Mimasu:2014nea,Aad:2015zva,Brivio:2017ije}. The bounds can be summarised as follows:
\be
\begin{aligned}
|g_{agg}|\lesssim&\,1.1\times 10^{-5}\GeV^{-1}\qquad
&&\text{for}\qquad m_a\,\,\lesssim60 \MeV
\\
|g_{agg}|\lesssim&\, 10^{-4}\GeV^{-1}\qquad
&&\text{for}\qquad 60 \MeV\lesssim m_a\,\,\lesssim0.1\GeV
\end{aligned}
\ee
that can be translated in terms of $f_a/|c_{agg}|$ as 
\be
\begin{aligned}
\dfrac{f_a}{|c_{agg}|}\gtrsim&\,2\times 10^{3}\GeV\qquad
&&\text{for}\qquad m_a\,\,\lesssim60 \MeV
\\
\dfrac{f_a}{|c_{agg}|}\gtrsim&\,2\times 10^{2}\GeV\qquad
&&\text{for}\qquad 60 \MeV\lesssim m_a\,\,\lesssim0.1\GeV
\end{aligned}
\ee
taking $\alpha_s(M_Z^2)=0.1184$.

The previous results holds for a stable or long lived axion, while if the lifetime is short enough $\Upsilon$ decays from BaBar data may provide interesting bounds: considering $\Upsilon(2s,3s)\to\gamma a(\to jj)$~\cite{Lees:2011wb} and $m_a=1\GeV$,
\be
\dfrac{f_a}{c_{agg}}\gtrsim80\GeV\,.
\ee
This bound is expected to be improved at Belle-II~\cite{CidVidal:2018blh} and reach values close to $0.2\TeV$.
\\

\noindent{\bf Couplings to massive gauge bosons}\\
\indent Rare meson decays provide strong constraints of axion couplings to two $W$ gauge bosons (as already discussed, no axion-SM fermion 
couplings are present at tree-level in the minimal AML$\s$M). The most relevant observable for axion masses below $\sim0.2\GeV$ is $K^+\to \pi^+a(a\to\text{inv.})$ whose branching ratio has been bounded by the E787 and E949 experiments~\cite{Adler:2008zza}:
\be
\mathcal{B}(K^+\to \pi^+a(a\to\text{inv.}))<7.3\times 10^{-11}\,.
\ee
For larger masses up to a few GeV's, the $B^+\to K^+a(a\to\text{inv.})$ decay provides the most stringent bound: BaBar experiment has proven that~\cite{Lees:2013kla}
\be
\mathcal{B}(B^+\to K^+a(a\to\text{inv.}))\lesssim 3.2\times 10^{-5}\,.
\ee
Belle-II expected sensitivity improves this bound of approximately one order of magnitude~\cite{Izaguirre:2016dfi,Cunliffe:2017cox}.

The induced bounds on $g_{aWW}$ effective coupling read~\cite{Izaguirre:2016dfi}:
\be
\begin{aligned}
|g_{aWW}|\lesssim&\,3\times 10^{-6}\GeV^{-1}\qquad
&&\text{for}\qquad m_a\,\,\lesssim 0.2 \GeV
\\
|g_{aWW}|\lesssim&\, 10^{-4}\GeV^{-1}\qquad
&&\text{for}\qquad 0.2 \GeV\lesssim m_a\,\,\lesssim 5\GeV
\end{aligned}
\ee
that can be translated in terms of $f_a/|c_{aWW}|$ as 
\be
\begin{aligned}
\dfrac{f_a}{|c_{aWW}|}\gtrsim&\,4\times 10^{2}\GeV\qquad
&&\text{for}\qquad m_a\,\,\lesssim0.2 \GeV
\\
\dfrac{f_a}{|c_{aWW}|}\gtrsim&\,10\GeV\qquad
&&\text{for}\qquad 0.2 \GeV\lesssim m_a\,\,\lesssim5\GeV\,.
\end{aligned}
\label{BaBarBound}
\ee

Collider searches are able to put independent constraints on $g_{aWW}$ as well as on couplings with other gauge bosons. Following 
Ref.~\cite{Brivio:2017ije}, considering LHC data with $\sqrt{s_\text{cm}}=13\TeV$ and for axion masses $m_a\lesssim1\GeV$, the mono-$W$, $pp\to a W(\to \mu\nu_\mu)$, and mono-$Z$, $pp\to aZ(\to ee)$, signals put the following constraints:
\be
|g_{aWW}|\lesssim1.6\times10^{-3}\GeV^{-1}\,,\qquad\qquad
|g_{aZZ}|\lesssim8\times10^{-4}\GeV^{-1}\,.
\ee
LEP data~\cite{Acciarri:1994gb,Anashkin:1999da}, on the other side, on the radiative $Z$ decays leads to a bound on $a\gamma Z$ coupling~\cite{Dolan:2017osp}:
\be
|g_{a\gamma Z}|\lesssim6.4\times 10^{-5}\GeV^{-1}\,.
\ee
The corresponding bounds on $f_a/|c_i|$ are given by:
\be
\dfrac{f_a}{|c_{aWW}|}\gtrsim0.7\GeV\,,\qquad
\dfrac{f_a}{|c_{aZZ}|}\gtrsim1.4\GeV\,,\qquad
\dfrac{f_a}{|c_{aZ\gamma}|}\gtrsim18\GeV\,.
\label{CouplingsFisicasRilevantes}
\ee
The future LHC sensitivity prospects for mono-$W$ and mono-$Z$ considering an integrated luminosity of $3000$ fb${^{-1}}$ improve the first two bounds of one order of magnitude~\cite{Brivio:2017ije}.

The previous limits from flavour factories and from colliders hold for an axion that escapes the detector and therefore is considered as missing energy in the data analyses. If instead the axion mass is large enough or its characteristic scale $f_a$ is low enough, the axion may decay within the detector and the previous limits cannot be taken into consideration. In these cases, the axion may decay into two photons, two gluons or two fermions, depending on its mass. 

Focussing on the radiative decay, LEP data~\cite{Acciarri:1994gb,Anashkin:1999da} on the decays $Z\to2\gamma$ and $Z\to3\gamma$ has been used to infer a bound on axion couplings in case of its decay inside the detector. From the analysis in Ref.~\cite{Dolan:2017osp} and assuming that $\mathcal{B}(a\to\gamma\gamma)=1$, the following constraints are found:
\be
\begin{aligned}
g_{a\gamma Z}\lesssim 6\times 10^{-4}\GeV^{-1}\qquad &\text{for}\qquad m_{\pi^0}\lesssim m_a\lesssim 10\GeV\\
g_{a\gamma Z}\lesssim 2\times 10^{-4}\GeV^{-1}\qquad &\text{for}\qquad 10\GeV\lesssim m_a\lesssim 91.2\GeV\,.
\end{aligned}
\ee
The corresponding bounds on $f_a/|c_i|$ read
\be
\begin{aligned}
\dfrac{f_a}{|c_{a\gamma Z}|}\gtrsim 1.8\GeV\qquad &\text{for}\qquad m_{\pi^0}\lesssim m_a\lesssim 10\GeV\\
\dfrac{f_a}{|c_{a\gamma Z}|}\gtrsim 5.4\GeV\qquad &\text{for}\qquad 10\GeV\lesssim m_a\lesssim 91.2\GeV\,.
\end{aligned}
\ee
Dedicated studies at LHC are expected to improve these bounds of two orders of magnitude~\cite{Dolan:2017osp}.

A future bound on $g_{aWW}$ will arise by studying the decay $B^\pm\to K^\pm\,a(\to2\gamma)$ at Belle-II~\cite{Izaguirre:2016dfi}. Assuming that a sensitivity of $10^{-6}$ will be reached on the branching ratio of $B\to K\gamma\gamma$, then limits on $g_{aWW}$ will improve of one order of magnitude: the strongest bound applies in the interval $0.4\GeV\lesssim m_a\lesssim 5\GeV$ and reads
\be
g_{aWW}\lesssim2\times 10^{-5}\GeV^{-1}\,,
\ee
that translates in terms of $f_a/|c_i|$ as
\be
\dfrac{f_a}{|c_{aWW}|}\gtrsim60\GeV\,.
\ee

All in all, these bounds with a decaying axion within the detector are very similar to the ones reported above in Eqs.~(\ref{BaBarBound}) and (\ref{CouplingsFisicasRilevantes}).
\\

\noindent{\bf The axion mass}\\
\indent There are two distinct contributions to the axion mass (gravitational and/or Planck-scale sources~\cite{Barr:1992qq,
Kamionkowski:1992mf,Holman:1992us,Alonso:2017avz} will not be discussed here). The first is due to purely QCD effects (axion mixing 
with neutral pions), which is estimated to be~\cite{Shifman:1979if,Bardeen:1978nq,DiVecchia:1980yfw}
\be
m_a\sim 6 \mueV\left(\dfrac{10^{12}\GeV}{f_a/c_{agg}}\right)\,,
\label{AxionMassContributionQCD}
\ee
for values of $f_a$ typically taken to be larger than $10^6\GeV$. The second is due to the extra fermions that couple to the axion, such as 
in the KSVZ invisible axion model~\cite{Kim:1979if,Shifman:1979if}:
\be
m_a=\dfrac{\sqrt{Z}}{1+Z}\dfrac{\alpha_s^2}{\pi^2}\dfrac{f_\pi}{f_a}m_\pi \ln\left(\dfrac{m_\psi^2}{m_u m_d}\right)\,,
\label{AxionMassContributionKSVZ}
\ee
where $Z\simeq m_u/m_d$ and $f_\pi\sim 94\MeV$ is the pion decay constant and $m_\psi$ is the generic mass of the exotic fermions. 
This contribution is a decreasing function with $f_a$ for values of $f_a>10\MeV$: considering similar values of $f_a$ and $m_\psi$, 
it follows that 
\be
\begin{aligned}
m_a&\sim 100 \keV\quad&&\text{for}\quad f_a\sim 1\GeV\\
m_a&\sim 0.2 \keV\quad&&\text{for}\quad f_a\sim 10^3\GeV\\
m_a&\sim 0.3 \eV\quad&&\text{for}\quad f_a\sim 10^6\GeV\\
m_a&\sim 0.004 \eV\quad&&\text{for}\quad f_a\sim 10^8\GeV\,.
\end{aligned}
\label{ExplicitAxionMasses}
\ee
Notice that, for the last two cases, the QCD mass in Eq.~(\ref{AxionMassContributionQCD}) is relevant and provides the dominant contributions 
of $60\eV$ and $0.6\eV$ respectively. These benchmarks are interesting for the discussion that follows.

\subsection{QCD Axion or Axion-Like-Particle?}

In Sect.~\ref{Sect.PhenoI}, three values for $f_s$ have been considered: $f_s=1\TeV$, $f_s=10^3\TeV$ and $f_s=10^6\TeV$. Eq.~(\ref{favsvr}) 
links the axion scale $f_a$ to the VEV of the radial component of $s$, and in consequence $f_a\simeq f_s$ in first approximation. The 
corresponding induced axion mass belongs to the window from tens of meV to the keV, according to Eq.~(\ref{ExplicitAxionMasses}). For 
this range of values, the strongest constraints on $f_a$ come from the axion coupling to two photons $g_{a\gamma\gamma}$, 
Eqs.~(\ref{Constraintsa2gamma}) and (\ref{Constraintsfaa2gamma}): specifying the value of $c_{a\gamma\gamma}$ for the minimal AML$\s$M 
charge assignment as reported in Tab.~\ref{tab:pheno}, one gets
\be
f_s\gtrsim 3.7\times 10^8\GeV\,.
\ee

It follows that a QCD axion, consistent with all the present data, can only be generated in the minimal AML$\s$M if the scale $f_s$, 
associated to the PQ breaking, is of the order of $10^8\GeV$ or larger. As discussed in Ref.~\cite{Brivio:2017sdm}, the resulting axion 
falls into the category of the so-called invisible axions~\cite{Kim:1979if,Shifman:1979if,Dine:1981rt,Zhitnitsky:1980tq}, as such a large 
$f_s$ scale strongly suppresses all the couplings with SM fermions and gauge bosons, preventing any possible detection at colliders or 
at low-energy (flavour) experiments. 

The difference with respect to the traditional invisible axion models resides partly in the axion couplings to photons and gluons, and 
in the EWSB sector. As underlined in Ref.~\cite{Brivio:2017sdm}, adding a KSVZ axion to the ML$\s$M narrows the range of possible values 
that the ratio $c_{a\gamma\gamma}/c_{agg}$ may take: the minimal AML$\s$M presented here provides a very sharp prediction for this ratio, 
\be
\dfrac{c_{a\gamma\gamma}}{c_{agg}}=\dfrac{14}{3}\,.
\ee
Moreover, in the minimal AML$\s$M with $f_s\gtrsim10^8\GeV$ the low-energy theory is not exactly the SM, but the EWSB mechanism is 
non-linearly realised and the Higgs particle originates as a GB. This model may be confirmed, or excluded, by a precise measure of 
$c_{a\gamma\gamma}/c_{agg}$ and by a dedicated analysis of the EW sector. In particular, this case corresponds to the scenario where only 
the physical Higgs remains in the low-energy spectrum, while the other two scalar dofs are very massive. In consequence, only indirect 
searches on Higgs couplings or the physics associated to the longitudinal components of the SM gauge bosons may have the potential to 
constrain the minimal AML$\s$M.\\

For much lighter values of the $f_s$ scale, instead, the astrophysical bounds on $g_{a\gamma\gamma}$ coupling can be satisfied only assuming that the axion mass and its characteristic scale $f_s$ are not correlated. This corresponds to the ALP scenario: differently from the QCD axion, an ALP has a mass that is independent from its characteristic scale $f_s$, due to additional sources of soft shift symmetry breaking with respect to those in Eqs.~(\ref{AxionMassContributionQCD}) and (\ref{AxionMassContributionKSVZ}), and does not necessarily solve the 
strong CP problem\footnote{In the ALP scenario, a solution to the Strong CP problem is not guaranteed and therefore the condition 4 is not required. An additional scenario satisfying conditions 1, 2, and 3, can be considered: in this case, $n_{q_L}=n_{\psi_L}=n_{\psi_R}=n_{\chi_R}=n_{t_R}\pm n_s=n_{\chi_L}\pm n_s$ (with the ``$+$'' or ``$-$'' are associated to the presence of the $z_1$ or $\tilde{z}_1$ terms in the Lagrangian, respectively), and the induced renormalisable scalar potential turns out to be the same as in Eq.~(\ref{renpot}).}. In what follows, an ALP with $m_a=1\GeV$ and $f_s\sim1\TeV$ will be considered as a benchmark point that passes the previous bounds on axion coupling to photons.

By increasing the axion mass and decreasing $f_s$, its decay length also decreases. The distance travelled by the axion after being produced may be casted as follows~\cite{Brivio:2017ije},
\be
d\approx\dfrac{10^4}{c_i^2}\left(\dfrac{\MeV}{m_a}\right)^4\left(\dfrac{f_s}{\GeV}\right)^2\left(\dfrac{|p_a|}{\GeV}\right)\m\,,
\ee
where $c_{i}$ are the couplings in Tab.~\ref{tab:pheno} and the typical momentum considered is $\gtrsim100\GeV$. 
The benchmark ALP decays predominantly into photons ($c_{a\gamma\gamma}\gg c_{agg}$), and its decay length is less than $1\mm$. The most sensitive observables to test this ALP will be $B^\pm\to K^\pm\, a(\to2\gamma)$ that could be investigate at Belle-II, and $Z\to3\gamma$ decay that could be studied at LHC: indeed, these processes would be sensitive to values of $f_a$ between $1\TeV$ and $6.5\TeV$.

\subsubsection*{The Fine-Tuning Problem}

The presence of different scales in the scalar potential leads to a fine-tuning problem in the model. As already mentioned, the parameter 
$\xi$ measures the tension between the EW scale and the $SO(5)$ SSB scale, as shown in Eq.~(\ref{C1expanded}). In models where axions or ALPs 
are dynamically originated, a new scale $f_s$ is present and typically much larger than the EW scale. Once the scalar field $s$ develops a VEV, 
the scale $f$ receives a contribution proportional to $\sqrt\lsp f_s$, as can be read in Eq.~(\ref{VSSB}). This leads to $f\approx f_s\gg v$, or  
$\lsp \ll 1$: this represents two sides of the same fine-tuning problem. This is actually the case for any QCD axion arising in the ML$\s$M. Instead, in the ALP model presented here $f_s\sim1\TeV$ and therefore no tuning is required on $\lsp$ (see also Ref.~\cite{Alonso-Gonzalez:2018vpc}).

\section{Concluding Remarks}
\label{Sect.Conclusions}

The AML$\s$M~\cite{Brivio:2017sdm} represents a class of models that extend the ML$\s$M~\cite{Feruglio:2016zvt} by the introduction of a complex 
scalar singlet, that allows to supplement the $SO(5)$ and EW symmetries with an extra $U(1)_\text{PQ}$.

The spectrum of the AML$\s$M encodes: i) the SM gauge bosons and fermions; ii) three real scalar dofs, one of them, the Higgs particle, being 
the only uneaten GB of the $SO(5)/SO(4)$ breaking; iii) two types of vectorial exotic fermions respectively in the fundamental and in the singlet 
representation of $SO(5)$; iv) the PQ GB originated by the spontaneous breaking of the $U(1)_\text{PQ}$ symmetry. The scale $f$ of the 
$SO(5)/SO(4)$ breaking is expected to be in the TeV region, in order to solve the Higgs hierarchy problem, while the PQ--breaking scale, $f_s$, 
is in principle independent from $f$, spanning over a large range of values.

A detailed analysis of the scalar potential and its minima has been presented for the first time. The appearance of possible $SO(5)$ and PQ 
explicit breaking terms arising from 1-loop fermionic and gauge contributions has been extensively discussed. The type and number of the 
additional terms required by renormalisability depends on the PQ charges assigned to the fields of the model.

A minimal AML$\s$M has been identified by introducing few general requirements with the intent to minimize the number of parameters in the 
whole Lagrangian. In particular, the parameter space of the minimal AML$\s$M scalar sector is determined by $7$ parameters. Two of them can 
be fixed by identifying one scalar dof with the physical Higgs particle and its VEV with the EW scale. The remaining free parameters correspond 
to: the quartic couplings $\lambda$ and $\lambda_s$ that control the linearity of the EWSB and the PQ symmetry breaking mechanisms, respectively; 
the scales $f$ and $f_s$ related to the symmetry breaking; the mixed quartic coupling $\lsp$ that represents the portal between the EW and 
PQ sectors. Simplified analytical expressions can be obtained for the scalar sector by integrating out the highest mass dof, either in the 
strongly interacting regime, $\lam_s \gg 1$, keeping free the scales $f_s$ and $f$ either in the perturbative regime, $\lam_s \lesssim 1$, 
but assuming instead a large hierarchy between the scales, $f_s \gg f$. Interesting analytical expression for the scalar sector in the regime 
$f_s \sim f$ can be obtained also in the limit $\beta, \lsp \ll 1$.

The analytical and numerical analysis of the parameter space points out that for $f,\,f_s\gtrsim1\TeV$ the heavier scalar dofs are unlikely 
to give signals at the present and future LHC run, while only the non-linearity of the EWSB mechanism would lead to interesting deviations 
from the SM predictions in Higgs and gauge boson sectors. 

The analysis of the PQ GB phenomenology reveals two possible scenarios: a light QCD axion or a heavy ALP. In the first case, the axion mass 
is expected in the range $[\text{meV},\,\text{keV}]$ and the strong bounds present on the axion coupling to two photons require that its 
characteristic scale $f_a\sim f_s$ must be larger than $10^5\TeV$, strongly suppressing all its interactions. This model represents a minimal 
invisible axion construction, where the EWSB mechanism is non-linearly realised and the physical Higgs particle arises as a GB. As can be 
realised from Eqs.~(\ref{renparameters})-(\ref{lambda&fRfs}), invisible axion models are, in general, strongly fine-tuned. In fact, the typical 
$SO(5)/SO(4)$ breaking scale of the effective theory obtained integrating out the heavy degrees of freedom ``naturally runs'' to the highest 
scale, $f_R \sim f_s$, reintroducing the EW hierarchy problem, $\xi \ll 1$. Alternatively, the tuning $\lsp=0$ can be introduced: this is, 
however, rather unnatural as no symmetry 
protects it.

In the second scenario, the ALP typically has a much larger mass, independent from the value of its characteristic scale. The benchmark $m_a=1\GeV$ and $f_s=1\TeV$ has been considered for concreteness as it passes the bounds on the axion coupling to photons. This ALP is likely to be detected both at LHC and at Belle-II, the best sensitivity being on the $aWW$ coupling. Moreover, values of $f_s$ close to $1\TeV$ do not introduce any fine-tuning on the model, contrary to what occurs in QCD axion models.

\section*{Acknowledgements}
The authors thank R. Alonso, F. Feruglio, P. Machado, and A. Nelson for useful discussions, and I. Brivio and B. Gavela for comments and 
suggestions on the preliminary version of the paper. L.M. thanks the department of Physics and Astronomy of the Universit\`a degli Studi 
di Padova and the Fermilab Theory Division for hospitality during the writing up of the paper. F.P. and S.R. thank the University of 
Washington for hospitality during the writing up of the paper.\\

The authors acknowledge partial financial support by the European Union's Horizon 2020 research and innovation programme under the Marie 
Sklodowska-Curie grant agreements No 690575 and No 674896. L.M. acknowledges partial financial support by the Spanish MINECO through the 
``Ram\'on y Cajal'' programme (RYC-2015-17173), and by the Spanish ``Agencia Estatal de Investigaci\'on'' (AEI) and the EU ``Fondo Europeo 
de Desarrollo Regional'' (FEDER) through the project FPA2016-78645-P, and through the Centro de excelencia Severo Ochoa Program under grant 
SEV-2016-0597.

\appendix

\section{The Generic Axion Lagrangian}
\label{App.GenericPQTransf}

The Lagrangian containing the axion couplings, in the basis where fermionic terms are shift-symmetry preserving, can be written as 
\begin{equation}
\mathscr{L}_a=
\dfrac{1}{2}\partial_\mu a\partial^\mu a+\Delta_\psi\dfrac{\partial_\mu a}{2f_a}\ov\psi\gamma^\mu\gamma^5\psi+\Delta_\chi\dfrac{\partial_\mu a}{2f_a}\ov\chi\gamma^\mu\gamma^5\chi+\Delta_{\psi'}\dfrac{\partial_\mu a}{2f_a}\ov{\psi'}\gamma^\mu\gamma^5\psi'+ \Delta_{\chi'}\dfrac{\partial_\mu a}{2f_a}\ov{\chi'}\gamma^\mu\gamma^5\chi'\,.
\ee
where $\Delta_f\equiv n_{f_L}-n_{f_R}$. Moreover, axion coupling to gauge bosons arise due to the anomalous nature of the PQ symmetry. The effective Lagrangian containing the contributions from all the fermions reads
\be
\begin{split}
\mathscr{L}^\text{eff}_{a}=
&-\dfrac{\alpha_s}{8\pi}\,\frac{a}{f_a}\sum \Big[5\left(\Delta_{\psi}+\Delta_{\psi'}\right)+\left(\Delta_{\chi}+\Delta_{\chi'}\right)\Big] G^a_{\mu\nu}\tilde{G}^{a\mu\nu}+\\
&-\dfrac{\alpha_2}{8\pi}\,\frac{a}{f_a}\sum6\left(\Delta_{\psi}+\Delta_{\psi'}\right)W_{\mu\nu}^a\tilde{W}^{a\mu\nu}+\\
&-\dfrac{\alpha_1}{8\pi}\,\frac{a}{f_a}\sum\Big[6\Delta_{\psi}\left(2Y_{X}^2+2Y_{Q}^2+Y_{T_5}^2\right)+6\Delta_{\chi}Y_{T_1}^2+\\
&\hspace{3cm}+6\Delta_{\psi'}\left(2Y_{X'}^2+2Y_{Q'}^2+Y_{B_5}^2\right)+6\Delta_{\chi'}Y_{B_1}^2\Big]B^{\mu\nu}\tilde{B}^{\mu\nu}\,,
\end{split}
\label{LSM_transformed}
\end{equation}
where $Y_i$ are the Hypercharges of the components of $\psi$ and $\chi$ (see Eq.~(\ref{PsiChiComponents})). The sum is meant over the different generations: in the specific setup considered here, it reduces to the third family only. 

Moving to the gauge boson physical basis, the axion couplings to the gauge field strengths are given by:
\begin{align}
\hspace{-1cm}
\mathscr{L}^\text{eff}_{a}=
&-\dfrac{\alpha_s}{8\pi}\,\frac{a}{f_a}\sum \Big[5\left(\Delta_{\psi}+\Delta_{\psi'}\right)+\left(\Delta_{\chi}+\Delta_{\chi'}\right)\Big] G^a_{\mu\nu}\tilde{G}^{a\mu\nu}+\nn\\
&-\dfrac{\alpha_{em}}{8\pi}\,\frac{a}{f_a}\sum\Big[6\Delta_{\psi}\left(1+ 2Y_{X}^2+2Y_{Q}^2+Y_{T_5}^2\right)+6\Delta_{\chi}Y_{T_1}^2+\nn\\
&\hspace{3cm}+6\Delta_{\psi'}\left(1+ 2Y_{X'}^2+2Y_{Q'}^2+Y_{B_5}^2\right)+6\Delta_{\chi_R}Y_{B_1}^2\Big]F_{\mu\nu}\tilde{F}^{\mu\nu}\,,\\
&-\dfrac{\alpha_{em}}{8\pi}\,\frac{a}{f_a}\sum\Bigg\{6\Delta_{\psi}\left[\dfrac{1}{\tan\theta_W^2}+ \tan^2\theta_W\left(2Y_{X}^2+2Y_{Q}^2+Y_{T_5}^2\right)\right]+6\Delta_{\chi}\tan^2\theta_W\, Y_{T_1}^2+\nn\\
&\hspace{3cm}+6\Delta_{\psi'}\left[\dfrac{1}{\tan^2\theta_W}+ \tan^2\theta_W\left( 2Y_{X'}^2+2Y_{Q'}^2+Y_{B_5}^2\right)\right]+6\Delta_{\chi'}\tan^2\theta_W\,Y_{B_1}^2\Bigg\}Z_{\mu\nu}\tilde{Z}^{\mu\nu}\,,\nn\\
&-\dfrac{\alpha_{em}}{8\pi}\,\frac{a}{f_a}\sum\Bigg\{12\Delta_{\psi}\left[\dfrac{1}{\tan\theta_W}- \tan\theta_W\left(2Y_{X}^2+2Y_{Q}^2+Y_{T_5}^2\right)\right]-12\Delta_{\chi}\tan\theta_W\, Y_{T_1}^2+\nn\\
&\hspace{3cm}+12\Delta_{\psi'}\left[\dfrac{1}{\tan\theta_W}- \tan\theta_W\left( 2Y_{X'}^2+2Y_{Q'}^2+Y_{B_5}^2\right)\right]-12\Delta_{\chi'}\tan\theta_W\,Y_{B_1}^2\Bigg\}F_{\mu\nu}\tilde{Z}^{\mu\nu}\,,\nn\\
&-\dfrac{\alpha_{em}}{8\pi}\,\frac{a}{f_a}\dfrac{12}{\sin^2\theta_W}\left(\Delta_{\psi}+\Delta_{\psi'}\right)W^+_{\mu\nu}\tilde{W}^{-\mu\nu}\nn
\end{align}
where $\theta_W$ is the Weinberg angle.


\footnotesize
%

\providecommand{\href}[2]{#2}\begingroup\raggedright\endgroup

\end{document}